\documentclass[fleqn,usenatbib,nofootinbib]{mnras}
\usepackage{graphicx}
\usepackage{amsmath,amssymb}
\usepackage{lastpage}
\usepackage[all]{hypcap}
\usepackage[T1]{fontenc}
\usepackage{float}
\usepackage{subfigure}
\usepackage{afterpage}
\usepackage[usenames]{color}
\usepackage{mathrsfs}
\usepackage{upgreek}
\usepackage{hyperref}
\usepackage{gensymb}
\usepackage{color}
\usepackage[dvipsnames]{xcolor}
\usepackage{longtable}
\usepackage{threeparttable}
\usepackage{multirow}
\usepackage{placeins}
\usepackage{bm}
\usepackage[capitalise]{cleveref}
\usepackage{enumitem}

\graphicspath{{Figures/}}

\newcommand{\go}{g_\text{obs}}
\newcommand{\gb}{g_\text{bar}}
\newcommand{\Ug}{\Upsilon_\text{gas}}
\newcommand{\Ud}{\Upsilon_\text{disk}}
\newcommand{\Ub}{\Upsilon_\text{bulge}}

\newcommand{\e}{e_\text{N}}
\newcommand{\ag}{\alpha_\text{grav}}
\newcommand{\sig}{\sigma_\text{int}}

\newcommand{\gN}{\ensuremath{g_\mathrm{N}}}
\newcommand{\eN}{\ensuremath{e_\mathrm{N}}}

\title[Tension in modified gravity MOND]{
On the tension between the Radial Acceleration Relation and Solar System quadrupole in modified gravity MOND
}
\author[Desmond, Hees \& Famaey]{
Harry~Desmond$^{1}$\thanks{harry.desmond@port.ac.uk}, Aur\'elien Hees$^{2}$\thanks{aurelien.hees@obspm.fr} and Benoit Famaey$^3$\thanks{benoit.famaey@astro.unistra.fr}\\
$^{1}$Institute of Cosmology \& Gravitation, University of Portsmouth, Dennis Sciama Building, Portsmouth, PO1 3FX, UK\\
$^{2}$SYRTE, Observatoire de Paris, Universit\'e PSL, CNRS, Sorbonne Universit\'e, LNE, 61 avenue de l'Observatoire 75014 Paris, France\\
$^{3}$Universit\'e de Strasbourg, CNRS  UMR  7550,  Observatoire  astronomique  de  Strasbourg, 67000  Strasbourg,  France\\
}

\pubyear{2023}

\begin{document}
\label{FirstPage}
\pagerange{\pageref{FirstPage}--\pageref{LastPage}}
\maketitle

\begin{abstract}
Modified Newtonian Dynamics (MOND), postulating a breakdown of Newtonian mechanics at low accelerations, has considerable success at explaining galaxy kinematics. However, the quadrupole of the gravitational field of the Solar System (SS) provides a strong constraint on the way in which Newtonian gravity can be modified. In this paper we assess the extent to which the AQUAL and QUMOND modified gravity formulations of MOND are capable of accounting simultaneously for the
Radial Acceleration Relation (RAR),
the Cassini measurement of the SS quadrupole and the kinematics of wide binaries in the Solar neighbourhood.
We achieve this by
inferring the location and sharpness of the MOND transition from the SPARC RAR under broad assumptions for the behaviour of the interpolating function and external field effect.
We constrain the same quantities from the
SS quadrupole,
finding that this requires a significantly sharper transition between the deep-MOND and Newtonian regimes than is allowed by the RAR (an $8.7\sigma$ tension under fiducial model assumptions). This may be relieved somewhat by allowing additional freedom in galaxies' mass-to-light ratios---which also improves the RAR fit---and more significantly (to $1.9\sigma$) by removing galaxies with bulges.
For the first time, we also apply to the SPARC RAR fit an AQUAL correction for flattened systems, obtaining similar results.
Finally we show that the SS quadrupole constraint implies, to high precision, no deviation from Newtonian gravity in nearby wide binaries, and speculate on possible resolutions of this tension between SS and galaxy data within the MOND paradigm.
\end{abstract}



\section{Introduction}\label{sec:intro}
The goal of the theory of gravity is to explain as many gravitational phenomena as possible with as few parameters. The theory of General Relativity (GR), based on Newtonian gravity, is successful on scales from sub-mm (from E\"{o}tv\"{o}s type experiments;~\citealt{Adelberger}) to tens of Gpc (the dynamics of the cosmos as a whole, assuming dark matter and dark energy). On galaxy scales the principal alternative is Modified Newtonian Dynamics (MOND;~\citealt{Milgrom_1,Milgrom_3,Milgrom_2}), which postulates a breakdown of Newtonian gravity or inertia at accelerations $\lesssim10^{-10}$ m s$^{-2}$. This has success at explaining several aspects of galaxy dynamics as well as a smattering of observations at other scales (see~\citealt{FamaeyMcG, Banik} for reviews).

MOND provides a framework within which to understand the otherwise surprising simplicity of galaxy dynamics. For example, the baryonic Tully--Fisher relation---the correlation between the asymptotic circular velocity and baryonic mass of rotation-supported galaxies---has very small intrinsic scatter and an almost perfect power-law shape across five orders of magnitude in mass~\citep{McGaugh_BTFR,Desmond_BTFR}. There are no residual correlations with galaxy size or baryonic surface density~\citep{FamaeyMcG,Federico_BTFR, Desmond_uncorrelated}, which one would typically expect to be produced by the dark matter halos of a $\Lambda$-Cold Dark Matter ($\Lambda$CDM) cosmology (based on GR;~\citealt{Desmond_TFR}). At the same time, galaxies of fixed baryonic mass, while sharing a similar asymptotic circular velocity, display rotation curves (RCs) with a vast diversity of inner shapes even in dwarf galaxies where, in $\Lambda$CDM, dark matter supplies most of the gravity. Indeed, it has long been known that the shapes of RCs \emph{do} depend strongly on baryonic surface density, even in dark matter-dominated galaxies~\citep[e.g.][]{zwaan:1995,deblok:1997,swaters:2009}. The RC shapes are thus not only diverse at a given mass-scale while sharing a single asymptotic velocity, but are uniform at a given baryonic surface density scale. In $\Lambda$CDM this implies a diversity of cored and cuspy dark matter profiles which is not currently understood~\citep{Oman_diversity,Ghari}. This may be explained by highly non-circular motions \citep{Roper_diversity}, at the price of producing distorted 2D velocity fields not resembling observed ones \citep[e.g.][]{kuzio:2011}.

All of this phenomenology for disc galaxies can actually be deduced from a single, apparently more fundamental, \emph{local} relation \citep{MLS,OneLaw} between the total gravitational acceleration at any point in the disc ($\go$) and that generated by the observed baryons ($\gb$). This Radial Acceleration Relation (RAR) has recently been shown to pass all posited tests for fundamentality, and appears to be the root cause of \emph{all} other radial dynamical correlations of disc galaxies~\citep{Stiskalek}.
For such systems, the RAR is nothing more than MOND written in terms of observables, so that, within that paradigm, all the above observations become readily understandable. There were already hints at the inception of MOND that the characteristic rotation velocities of galaxies were correlated with their luminosity~\citep{TFR};~\citet{Milgrom_1,Milgrom_3,Milgrom_2} promoted them to a tentative law of Nature by proposing that Newtonian dynamics, the weak-field limit of GR, breaks down at ultra-low accelerations typical of the outskirts of galaxies (below a new universal constant $a_0 \simeq 10^{-10}$~m/s$^2$) where one instead has
$\go = \sqrt{a_0 \: \gb}$.
In semi-analytic and hydrodynamical models of galaxy formation of $\Lambda$CDM, on the other hand, while the global shape of the RAR can be reproduced, its low scatter and lack of residual correlations are very difficult to understand (e.g.~\citealt{DC_Lelli,Desmond_MDAR,Paranjape_Sheth}).

Given MOND's success at the scale of galaxies, it is crucial to investigate the extent to which it may be extended to other regimes. On larger scales, it has problems accounting for the mass discrepancy in clusters (e.g.~\citealt{clusters_1,clusters_2,clusters_3,clusters_4,Li_clusters}) and a fully-fledged and observationally consistent cosmology is proving challenging to formulate (although see \citealt{skordis} for recent progress). On smaller scales, within the Milky Way, key tests are the orbits of wide binary (WB) systems in which the internal orbital acceleration lies below $a_0$, and of bodies in the Solar System (SS). Both of these systems are embedded within the gravitational field of the Milky Way at or near the position of the Sun, which is around $1.8 \: a_0$~\citep{klioner:2021}. This makes the tests less clean than the fully weak-field probes in the outskirts of low surface density galaxies, but it is nevertheless clear---at least within canonical modified gravity formulations---that MOND should provide a small boost to the dynamics in these systems if it is to provide one for RCs at accelerations of $1.8 \: a_0$.
WBs have had a long and confused history in the context of MOND, with some authors
finding consistency with Newtonian dynamics~\citep{Pittordis_Sutherland, Banik_WBT}, while others argue that these analyses are flawed and the data instead prefer MOND~\citep{Hernandez, Hernandez_2, Hernandez_3, Hernandez_4, chae_wb_2, chae_wb, Hernandez_Chae}.

In the present work, we focus instead on the SS tests, which were revolutionised by the Cassini spacecraft's radioscience tracking~\citep{wolf:1995ys,antreasian:2005kx,cassini_1,iess:2010ys,iess:2012vn}. Cassini's measurements made it possible to reconstruct the spacecraft's trajectory as it orbited Saturn between 2004 and 2013, and to improve significantly our knowledge of Saturn's orbit. The most constraining measurement in the MOND context is a null detection of the quadrupole moment of the gravitational field of the Sun~\citep{hees:2014jk}. A quadrupole is expected in (at least the most popular) modified gravity formulations of MOND---the AQUAdratic Lagrangian (AQUAL;~\citealt{aqual}) and QUasilinear MOND (QUMOND;~\citealt{qumond}) formulations---given the external field effect (EFE) due to the relative magnitude and orientation of the (internal) Saturn--Sun and (external) Sun--Galactic Centre gravitational fields ~\citep{milgrom:2009vn,blanchet:2011ys}. \citet{Hees_main} found Cassini's quadrupole measurement to be inconsistent with a large variety of Newton-to-deep-MOND transitions that describe the RAR well.
Although the Cassini mission ended in 2017, studies of RCs have greatly advanced since then, making the topic ripe for revisiting.

The EFE arises because, as an acceleration-based modification to Newtonian mechanics, MOND is sensitive to the \emph{total} gravitational field and hence violates the strong equivalence principle \citep[see e.g.][for a detailed account]{FamaeyMcG, Banik}. While the majority of evidence seems to point towards the helpfulness of the EFE for explaining the kinematics and tidal stability of various types of galaxies within a MOND context~\citep[e.g.,][]{Famaey_2007a,Wu_2008,Haghi,McGaugh_Milgrom,Hees_main,Haghi_2016,Famaey_2018,Kroupa_2018,Thomas,Haghi_2019,Paper_I,Banik_M33,Oria,Kroupa_2022,Asencio}, there are also cases where the EFE should be present but appears absent~\citep{Freundlich}. It is also important to note that, while some form of EFE is generically expected by MOND, in modified inertia formulations it may be very different (e.g. dependent on an object's entire past trajectory) or effectively absent~\citep[e.g.][]{Milgrom_2011,MI_2023}.

The consistency---or lack thereof---between the RAR and SS quadrupole in MOND hinges on the nature of the transition between the Newtonian and deep-MOND regimes, $\go=f(\gb)$ in the language of the RAR. The functional form cannot, at present, be derived theoretically and thus must be constrained empirically: the only requirements in MOND (absent the EFE) are the limits $f(\gb)\rightarrow\gb$ as $\gb\rightarrow\infty$ and $f(\gb)\rightarrow\sqrt{a_0 \: \gb}$ as $\gb\rightarrow0$. Any $f$ that satisfies these limits may be re-expressed as a MONDian ``interpolating function'' (IF). Such transition functions are not part of MOND's basic tenets but are used, in one way or another, by all known modified gravity formulations of the theory, generally put by hand into the Lagrangian. The IF could however emerge from a fundamental underlying theory with an actually different functional form in different systems, as may in fact be expected in modified inertia formulations \citep{MI_2023}.

Many functions satisfying the MOND requirements have been proposed and investigated in the literature (see~\citealt{FamaeyMcG,Banik} and references therein). \citet{ESR-RAR} evaluated all possible functions of low complexity on dynamical galaxy data, concluding that the optimal function for RC data is fairly complex---and
may not have MOND limits---but that more data is required to deduce it unambiguously. In particular, it is not clear from RC data alone that $\go \propto \sqrt{\gb}$ at low $\gb$. Additional uncertainty arises from the EFE, which depends on the imperfectly observationally characterised baryonic large-scale structure of the Universe and has a functional form that cannot be deduced analytically for general mass distributions.
Here we group IFs into classes or parametric ``families'', which contain not only $a_0$ as a degree of freedom but also a {\it shape} parameter controlling the sharpness of the transition from the deep-MOND to Newtonian regimes.

The aim of this paper is twofold. First, we extend the RAR analysis of \citet{uRAR} to cover three IF families under four different state-of-the-art models for the EFE. Within classical modified gravity formulations (AQUAL and QUMOND), this essentially fully spans the space of possibilities both for the form of the MOND force law and the effect of the environment on the galaxies in question, thus affording a definitive inference of the associated parameters
with the galaxies' systematically uncertain properties fully accounted for. We also explore various priors for the mass-to-light ratios of the galaxies' components, in case the fiducial model is overly restrictive, and we explicitly check the effect of the flat disc geometry in the context of AQUAL. Second, we compare the RAR $\{a_0, \text{shape}\}$ constraints with those inferred from the measurement of the SS quadrupole by the Cassini mission, and to
results of the~\citet{Banik_WBT} wide binary test (WBT). These all provide independent probes of MOND, which, if inconsistent with each other, would pose a severe problem for the AQUAL~\citep{aqual} and QUMOND~\citep{qumond} weak-field modified gravity formulations.

In Sec.~\ref{sec:data} we describe the galaxy and SS data that we use as constraints. In Sec.~\ref{sec:method} we lay out our MOND models (Sec.~\ref{sec:rar_functions}), our method for inferring their parameters from the RAR (Sec.~\ref{sec:rar_inference}) and SS quadrupole (Sec.~\ref{sec:quadrupole_inference}), and the method for relating these to the WBT (Sec.~\ref{sec:WBT_method}). Sec.~\ref{sec:results} describes the results, separately for the RAR (Sec.~\ref{sec:results_rar}) and SS quadrupole (Sec.~\ref{sec:results_q2}), and explores why the two sets of constraints are difficult to reconcile~(Sec.~\ref{sec:new_if}).
We conclude in Sec.~\ref{sec:conc}. Throughout the paper, $\log$ has base 10.

\section{Observational data}
\label{sec:data}

\subsection{SPARC galaxy sample}
\label{sec:data_sparc}

For the RAR we utilise the \textit{Spitzer Photometry and Accurate Rotation Curves (SPARC)} sample~\citep{SPARC},\footnote{\url{http://astroweb.cwru.edu/SPARC/}} containing 175 RCs from the literature with photometry at 3.6$\mu$m from the \textit{Spitzer} satellite \citep[see also][]{Gentile}. We adopt the same quality cuts as~\citet{OneLaw}, excluding galaxies with quality flag 3 (indicating strong asymmetries, non-circular motions and/or offsets between the stellar and H\textsc{i} distributions) or inclination $i < 30^\circ$, and points with a fractional rotation velocity uncertainty $>10$ per cent. 2696 points from 147 galaxies remain, of which 31 contain a central stellar bulge. This is the same sample as was used in \citet{uRAR}.

\subsection{Cassini measurement}
\label{sec:data_cassini}

The MOND paradigm breaks the strong equivalence principle such that the external gravitational field of our Galaxy impacts the internal dynamics of the SS \citep{milgrom:2009vn,blanchet:2011ys}. The leading effect is a modification to the central Newtonian potential of the Sun by an additional quadrupole term (adopting the Einstein summation convention)
\begin{equation}\label{eq:SS_potential}
    \delta \Phi(\bm x) = -\frac{Q_2}{2} x^i x^j\left(\hat{e}_i \hat{e}_j - \frac{1}{3} \delta_{ij}\right) \, ,
\end{equation}
where ${\bm \hat{\bm e}}= {\bm g}_{\rm ext}/g_{\rm ext}$ is a unit vector pointing towards the Galactic Centre, $\bm x$ the position within the SS with respect to the Sun, $\delta_{ij}$ the Kronecker delta and $Q_2$ a parameter that depends on the MOND IF and acceleration scale $a_0$, the Sun's mass $M$ and the value of the external gravitational field from the Galaxy, $g_{\rm ext}$.

The modified Newtonian potential from Eq.~\ref{eq:SS_potential} will induce an anomalous acceleration which rises linearly with distance from the Sun and can be sought using the kinematics of planets \citep{milgrom:2009vn,blanchet:2011ys, hees:2012fk}, Kuiper Belt objects \citep{Brown}, asteroids or comets \citep{maquet:2015jk, 2024:vokrouhlicky}. In \cite{hees:2014jk}, the $Q_2$ parameter of Eq.~\ref{eq:SS_potential} was inferred using the DE430 planetary ephemerides data \citep{folkner:2014uq}. Of central importance for constraining this parameter are the 9 years of Cassini range and Doppler tracking data which strongly constrain Saturn's orbit. Although subject to some systematic uncertainty (see \citealt{hees:2014jk} for details), the $1\sigma$ constraint may be described by
\begin{equation}\label{eq:Q2_measurement}
    Q_2 = \left(3\pm 3\right) \times 10^{-27} \: \mathrm{s}^{-2} \, .
\end{equation}

\section{Method}
\label{sec:method}

\subsection{Parametrising MOND}
\label{sec:rar_functions}

We consider a diverse set of possible MONDian descriptions of the relation between total and baryonic acceleration, viz. IFs.
The MOND force law in highly symmetric configurations can be expressed as:
\begin{equation}\label{eq:g}
    \bm g = \bm \gN  \ \nu\left(\frac{\gN}{a_0}\right) \, ,
\end{equation}
where $\bm g$ is the total gravitational field (or dynamical acceleration), $\bm \gN$ is the Newtonian gravitational field sourced by baryons, $\nu(y)$ is the IF and $\gN\equiv\left|\bm \gN\right|$. The only {\it a priori} requirement on $\nu$ from the basic tenets of MOND is the asymptotic behaviour $\nu(y) \rightarrow 1$ when $y \rightarrow \infty$, and $\nu(y) \rightarrow 1/\sqrt{y}$ when $y \rightarrow 0$. The three most common examples are the ``Simple'' IF~\citep{Simple_IF,zhao:2006la}
\begin{equation}
    \nu_\text{simp}(y) = \frac{1 + (1 + 4 y^{-1})^{1/2}}{2}\:,
\end{equation}
the ``Standard'' IF \citep{Milgrom_2}:
\begin{equation}
    \nu_\text{stand}(y) = \left(\frac{1 + (1 + 4 y^{-2})^{1/2}}{2}\right)^{1/2}
\end{equation}
and the ``RAR'' (also called ``McGaugh--Lelli--Schombert'' or ``MLS'') IF~\citep{MLS}
\begin{equation}
\label{eq:nurar}
    \nu_{\rm RAR}(y) = \left(1 - {\rm exp}(-y^{1/2})\right)^{-1}.
\end{equation}

Absent theoretical or compelling observational evidence for a particular IF, it is convenient to group IFs into ``families'' sharing a more general functional form in order to test them. These have a parameter (in addition to $\gN/a_0$, their only other degree of freedom) that interpolates between specific IFs by controlling the sharpness of the transition between the deep-MOND ($\gN/a_0\ll1$) and Newtonian ($\gN/a_0\gg1$) regimes.
Here, we investigate three IF families described in \citet{FamaeyMcG}:
\begin{subequations}\label{eq:if_fams}
	\begin{eqnarray}
    \nu_n(x)&=&\left[\frac{1+\left(1+4x^{-n}\right)^{1/2}}{2}\right]^{1/n}\, ,  \label{eq:nun}\\
    \nu_\delta(x)&=&\left(1-e^{-x^{\delta/2}}\right)^{-1/\delta}\, , \label{eq:IF_delta}\\
	\nu_\gamma(x)&=&\left(1-e^{-x^{\gamma/2}}\right)^{-1/\gamma}+\left(1-1/\gamma\right) e^{-x^{\gamma/2}} \, . \label{eq:IF_gamma}
	\end{eqnarray}
\end{subequations}
These cover the great majority of functions that have been considered in the literature;
in particular, the $n$-family encompasses the Simple ($n=1$) and Standard ($n=2$) IFs, while the $\delta$- and $\gamma$-families contain the RAR IF at $\delta=\gamma=1$.\footnote{There is also a $\beta$ family (eq. 51 of \citealt{FamaeyMcG}), but this is not flexible enough to cover Simple- or RAR-like behaviour, and hence is poor at fitting the RAR.} We refer to $n$, $\delta$ and $\gamma$ collectively as ``shape''.

As well as a model without any EFE, we explore the possible consequences of the EFE on the RAR fit using two different formulae.
These are the ``Freundlich--Oria analytic''~\citep{Freundlich,Oria} and ``AQUAL numerical'' models of \citet{chae:2022tt} (cases 4 and 6 in their table 1). These are both fitting functions for RCs, the former developed in the context of the QUMOND model and the latter that of AQUAL. These modify the raw IFs and hence can be applied to any of them. Considering the Newtonian external field strength in units of $a_0$, i.e. $e_{\rm N} \equiv g_{\rm N, ext}/a_0$, the QUMOND EFE model is given by
\begin{equation}\label{eq:efe_fo}
\nu_\text{EFE, QUMOND}(y) = \nu\left({\rm min}\left[y+\frac{e_{\rm N}^2}{3y},e_{\rm N}+\frac{y^2}{3e_{\rm N}}\right]\right)
\end{equation}
and the AQUAL model by
\begin{equation}\label{eq:efe_an}
\nu_\text{EFE, AQUAL}(y) = \nu(y_\beta) \left[1+\tanh\left(\frac{\beta e_{\rm N}}{y}\right)^{\gamma} \frac{\hat{\nu}(y_\beta)}{3} \right],
\end{equation}
where $\hat{\nu}(y)\equiv \text{d}\ln \nu(y)/\text{d}\ln y$ and $y_\beta \equiv \sqrt{y^2+(\beta e_{\rm N})^2}$. The best-fit values of the parameters $\beta$ and $\gamma$ are 1.1 and 1.2 respectively~\citep{chae:2022tt}. These roughly span the space of possible EFE behaviour, although note that Eq.~\ref{eq:efe_an} is strictly only intended to be used in the outer parts of RCs (where the EFE plays a larger role).

For each of these models we then consider two different ways of setting the external field strengths $\eN$ of the SPARC galaxies. The first is simply to treat $\eN$ as a {\it global} constant and infer it (with a wide uniform prior) from the data. The second is to allow the {\it local} $\eN$ to vary galaxy-by-galaxy, with a prior specified by the ``maximum clustering'' model of \citet{Chae_2}, which infers $\eN$ for each SPARC galaxy from its large-scale baryonic environment assuming unseen baryons to correlate maximally with observed baryons.\footnote{\citet{Chae_2} also obtain ``no clustering'' results in which unseen baryons do not correlate at all with observed baryons and hence do not systematically increase $\eN$ but only its scatter, and \citet{uRAR} considers an ``average clustering'' model midway between the two. We focus on the maximum-clustering case because it was found to yield best agreement with the data in \citet{Chae_2,Chae_3}, and is \textit{a priori} the most likely case in MOND. The differences for our purposes are minor.} $\eN$ for each galaxy separately is then a free parameter of the inference to be marginalised over. As in \citet{uRAR}, for galaxies outside the footprint of the Sloan Digital Sky Survey (where \citealt{Chae_2} could not calculate precise $\eN$ values), we use the median over all SPARC galaxies within the footprint as the prior centre, with an uncertainty twice the median uncertainty for those galaxies. For the maximum-clustering case this gives the prior $\log(\eN) = -2.300 \pm 0.575$.

Finally, it is important to note that in MOND Eq.~\ref{eq:g} (often referred to as the ``algebraic MOND relation'') is exact only for circular orbits in modified inertia versions of the paradigm \citep{MI_1994, inertia_1}, and cannot be strictly exact in modified gravity, even in the absence of an EFE. We will consider it a sufficient approximation~\citep{Jones-Smith,Oria} in our fiducial analysis, but in Sec.~\ref{sec:results_MG} we will investigate an alternative formula for the MOND boost taking into account the correction for flattened systems within the AQUAL formulation.

\subsection{RAR inference}
\label{sec:rar_inference}

Our method to analyse the RAR extends that of \citet[][see especially sec.~3.2]{uRAR}. In short, we infer the parameters of the RAR ($a_0$ and intrinsic scatter $\sig$, in addition to the IF shape) simultaneously with the parameters describing each galaxy using priors from previous measurements and theoretical expectations. The galaxy nuisance parameters are distance $D$, inclination $i$, luminosity $L_{3.6}$, the mass-to-light ratios ($M/L$) of the disc $\Ud$, bulge $\Ub$ and gas $\Ug$, and the external field strengths $\eN$. This amounts to $\sim$900 parameters in total, which is too high a dimensionality for most Markov Chain Monte Carlo (MCMC) samplers to handle reliably. It is however necessary in order to properly propagate the galaxies' parameters as systematic rather than statistical uncertainties, and map out their degeneracies with the global properties of the RAR. Leveraging automatic differentiation in Jax, we employ the No U-Turns sampler (NUTS, a species of Hamiltonian Monte Carlo) as implemented in \texttt{numpyro}~\citep{phan2019composable, bingham2019pyro}. This produces a fully converged chain of $\sim$2000 points in $\sim$10 minutes; we concatenate 28 chains with multiprocessing for improved statistics and to verify insensitivity to the initialisation. Restricting to the Simple IF and a single parametrisation of the EFE, \citet{uRAR} revealed the intrinsic scatter of the RAR to be minute ($\sig=0.034 \pm 0.001 \text{(stat)} \pm 0.001 \text{(sys) dex}$), supporting the claim that the RAR is ``tantamount to a law of nature''~\citep{OneLaw}. Weak evidence was adduced for the EFE.

Besides freeing up the IF and EFE implementations as discussed above,
we now develop more flexible models for the mass-to-light ratios of the SPARC galaxies' disks and bulges. These are the only parameters of the galaxies that are not directly constrained empirically, yet are crucial for locating the galaxies' RC points on the RAR plane and hence constraining the RAR parameters. In particular we consider six possible priors:
\begin{enumerate}
\item The fiducial SPARC model in which $\Ud$ and $\Ub$ follow lognormal\footnote{We use lognormal for consistency with previous works, but the results with a normal prior are almost identical.} priors with means 0.5 and 0.7 and widths 0.125 and 0.175 respectively (this is also the model used in \citealt{uRAR}).
\item $\Ud$ and $\Ub$ drawn from separate Gaussian hyperpriors with means $\mu_\text{d}$ and $\mu_\text{b}$ inferred from the data with independent wide uniform priors. We retain a width of 25 per cent for the hyperprior. This models a scenario in which the centres of the prior $M/L$ distributions are unknown, although their uncertainties follow the SPARC error model.
\item As (ii) but requiring $\mu_\text{b}>\mu_\text{d}$, as expected from population synthesis models.
\item $\Ud$ and $\Ub$ drawn independently from wide uniform priors, modelling a  scenario in which nothing is known about them \textit{a priori}.
\item As (iv) but requiring $\Ub>\Ud$ on a galaxy-by-galaxy basis, as expected from population synthesis models.
\item As (ii) but removing the 31 galaxies with bulges. This models a scenario in which bulges are too poorly understood to be included in the sample.
\item As (ii) but removing the 116 galaxies without bulges. This provides a counterpoint to (vi), allowing us to investigate the consistency of bulgey and bulge-free galaxies.\footnote{Very similar results for models (vi) and (vii) are obtained using free uniform priors rather than Gaussian hyperpriors for the remaining $\Upsilon$ parameters.}
\end{enumerate}
The $\Ug$ model, based on \citet{research_note} with a 10 per cent uncertainty, is not altered. Adding extra dark gas has been shown not to improve MOND RC fits~\citep{gas_scaling}.

We use uniform priors on $a_0$, $\sig$ and shape, all of which are well enough constrained by the data for the choice of prior to be unimportant. In particular, using a log-prior on the dimensionful quantities $a_0$ or $\sig$, or a prior on shape corresponding to a flat prior on $Q_2$ (see Sec.~\ref{sec:quadrupole_inference}) makes negligible difference to the RAR results.

\subsection{Quadrupole inference}
\label{sec:quadrupole_inference}

Theoretically, the SS quadrupole value that appears in Eq.~\ref{eq:SS_potential}  can be expressed as \citep{milgrom:2009vn}
\begin{equation}\label{eq:Q2}
Q_2 = -\frac{3a_0^{3/2}}{2\sqrt{GM}}q(\tilde{e})\, ,
\end{equation}
where, in the SS, $M=1 \, {\rm M}_\odot$, and $q$ is a dimensionless parameter that depends only on the MOND IF as well as the value of
\begin{equation}
\tilde{e} \equiv \frac{g_{\rm ext}}{a_0},
\end{equation}
where $g_{\rm ext}$ in this case is the external field of the Milky Way at the SS.
In the context of QUMOND, \cite{milgrom:2009vn} has derived an exact expression for $q$:
\begin{equation}\label{eq:q}
    q(\tilde{e})=\frac{3}{2}\int_0^\infty \mathrm{d} v \int_{-1}^1 d\xi \left(\nu -1\right)\Big[e_{\rm N}\left(3\xi -5\xi^3\right)+v^2\left(1-3\xi^2\right)\Big] \, ,
\end{equation}
where $\nu=\nu\left[\sqrt{e_{\rm N}^2+v^4+2e_{\rm N}v^2\xi}\right]$ and $e_{\rm N}$ is the solution of $e_{\rm N}\nu\left(e_{\rm N}\right)=\tilde{e}$. In the case of AQUAL, the above integral leads only to an approximate value for the $q$ parameter. In this case, $q$ must be computed by numerically solving the non-linear Poisson equation (as done in \citealt{milgrom:2009vn} and \citealt{blanchet:2011ys}). By comparing the QUMOND values with those obtained using AQUAL in table~1 of \cite{milgrom:2009vn} and table~1 of \cite{blanchet:2011ys}, it is clear that our QUMOND calculation leads to a lower $Q_2$,\footnote{More precisely, we compare the values of $Q_2$ obtained for $\nu_n$ with $n=1,2,5$ and $20$. The value of $Q_2$ is always larger in AQUAL, by less than $10^{-27}$ s$^{-2}$, except for $n=1$ where it is about 25\% larger. The tension is therefore if anything underestimated in our QUMOND calculation.} and is therefore conservative when it comes to assessing the tension with the data.

\begin{figure}
  \centering
  \includegraphics[width=0.48\textwidth]{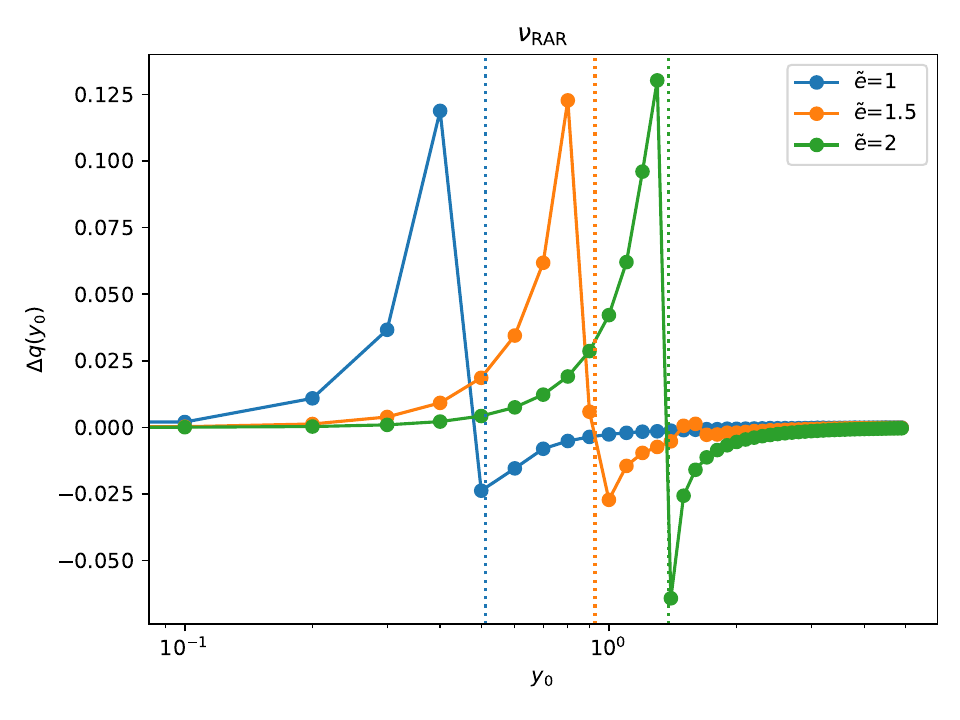}
  \caption{Variation of $\Delta q(y_0)$, the contribution from $\nu(y)$ between $y_0$ and $y_0+0.1$ to $q(\tilde e)$, for the IF $\nu_{\rm RAR}$ of Eq.~\ref{eq:nurar} and three different values of $\tilde e$. The value of $q(\tilde e)$ from Eq.~\ref{eq:q} equals the sum of the points for each curve, which gives $q(1)=0.094$, $q(1.5)=0.159$ and $q(2)=0.221$. The vertical dashed lines represent the corresponding values of $e_N$. The plot shows that the value of $q(\tilde e)$ mainly probes the behaviour of the IF around $e_N$.}
  \label{fig:Delta_q}
\end{figure}

The value of $q(\tilde e)$ obtained from Eq.~\ref{eq:q} depends mainly on the behaviour of the IF around $e_N$. To illustrate this, Fig.~\ref{fig:Delta_q} presents, as a function of the argument of the $\nu$-function in a small bin between $y_0$ and $y_0+\Delta y$ (with $\Delta y$ set to 0.1), the contribution $\Delta q(y_0)$ of $\nu(y)$ in this bin to the value of $q(\tilde e)$. In practice, $\Delta q(y_0)$ is computed by replacing the function $\nu(y)$ by 1 in Eq.~\ref{eq:q} (i.e. $\bm{g} = \bm{g_\text{N}}$) except in the range $[y_0, y_0+\Delta y]$ where it instead follows Eq.~\ref{eq:nurar} (i.e., $\delta=1$ in Eq.~\ref{eq:IF_delta} or $\gamma=1$ in Eq.~\ref{eq:IF_gamma}). Fig.~\ref{fig:Delta_q} presents the variation of $\Delta q$ with $y_0$ for three different values of the external field $\tilde e$. The vertical dashed lines show the corresponding values of $e_N$. It can be noticed that $\left|\Delta q\right|$ is large only around $e_N$, which is the case for all other IFs too. This means that the Cassini constraint probes the IF around $y\sim e_{\rm N}=g_\mathrm{N,ext}/a_0$, the gravitational field of the Milky Way at the Sun, similarly to the WBT \citep{Banik_WBT} and to the analysis of \citet[][]{2024:vokrouhlicky}. This is independent of whether the internal Newtonian acceleration probed is higher (as in the Cassini constraint) or lower (as in the aphelia of long period comets) than $g_\mathrm{N,ext}$ in these different cases.

Then, for each IF family (Eq.~\ref{eq:if_fams}), we have computed the $q$ factor on a regular grid for the  $\left(\tilde{e}, \mathrm{shape}\right)$ parameters with $\tilde{e}$ ranging between 1 and 20 (with spacing 0.05) and the shape parameter between 0.5 and 10.4 (with spacing 0.1). To reach lower values of $a_0$, we supplement this with an irregular sampling of $\tilde{e}$ values extending to 1000. We then interpolate this grid using a 2D cubic spline.
We use Eq.~\ref{eq:Q2_measurement} as the likelihood function in our inference to constrain \{$a_0$, shape\}.  Our model for the external field of the SS sourced by the Milky Way derives from the Gaia EDR3 measurement of the acceleration of the Sun, which is $g_\text{ext}=2.32\pm0.16\times10^{-10}$ m s$^{-2}$ \citep{klioner:2021}. While the $2\sigma$ lower bound $2\times10^{-10}$ m s$^{-2}$ is realistic (corresponding to a large but acceptable peculiar velocity of the Sun;~\citealt{Bovy}), the 2$\sigma$ upper bound is not (as it would imply an unrealistic negative peculiar velocity of the Sun). We therefore consider the upper bound on $g_{\rm ext}$ to be $2.48\times\:10^{-10}$m s$^{-2}$. We treat it in one of two ways: i) use the value within the range [$2, 2.48$]$\times10^{-10}$ m s$^{-2}$ that leads to the lowest predicted value of $Q_2$ in order to be conservative in our inference of \{$a_0$, shape\} and maximise consistency with the RAR, or ii) infer $g_{\rm ext}$ along with \{$a_0$, shape\} using the prior $\mathcal{N}(2.32,0.16)$ truncated at 2.48. These methods give very similar results; we present results for the second method because it is more statistically principled and makes the $Q_2$ and RAR inferences fully independent.

For the $Q_2$ inference, a flat prior on shape produces a posterior extending to infinity, in which limit $Q_2\rightarrow0$. As well as complicating the construction of confidence intervals, this would introduce a strong volume effect into the posterior predictive distribution (PPD) of the WBT test statistic (see next subsection).
We therefore instead adopt a prior flat on $Q_2$ (with a uniform prior on $a_0$), which matches that used in \citet{hees:2014jk} to derive the constraint of Eq.~\ref{eq:Q2_measurement}. This is achieved by taking the numerical derivative of $Q_2$ with respect to shape at each point within the grid.

\subsection{Connection to the wide binary test}
\label{sec:WBT_method}

Following~\citet{Banik_WBT}, we summarise the expected result of a WBT in the Solar neighbourhood using a parameter $\ag$ defined as
\begin{equation}\label{eq:ag}
\ag \equiv \frac{\sqrt{\eta}-1}{0.193},
\end{equation}
where $\eta \equiv \langle g_r \rangle / g_\text{N}$, the boost of the
azimuthally averaged asymptotic radial gravity compared to Newton. In QUMOND this is given by
\begin{equation}
\eta = \nu_e\left(1+\frac{1}{3} \frac{\partial\ln\nu_e}{\partial\ln(g_\text{N,ext})}\right),
\end{equation}
where $\nu_e\equiv\nu\left(e_{\rm N}\right)$ is the IF evaluated at the Newtonian-equivalent Galactic gravity $e_{\rm N} = g_\text{N,ext}/a_0$ at the Solar position. In AQUAL it is instead given by
\begin{equation}
\eta = \mu_e^{-1} \left(\frac{\partial\ln\mu_e}{\partial\ln g_{\rm ext}}\right)^{-1/2} \times \tan^{-1}\left(\left(\frac{\partial\ln\mu_e}{\partial\ln g_{\rm ext}}\right)^{1/2}\right).
\end{equation}
$\mu$ is an alternative description of the IF, defined by $\mu(x)=\nu(y)^{-1}$ where $y=x\mu(x)$.
Eq.~\ref{eq:ag} is chosen such that $\ag=0$ for fully Newtonian gravity, and $\ag=1$ for the case of QUMOND with the Simple IF, assuming $a_0=1.2\times10^{-10}$ m s$^{-2}$ and $\tilde{e} = 1.8$. Table~4 of~\citet{Banik_WBT} shows the $\ag$ values corresponding to a few choices of IF at fixed $a_0$ and $g_\text{N,ext}$.

Eq.~\ref{eq:ag} describes a mapping from \{$a_0$, shape, $g_\text{N,ext}$\}
to $\ag$, allowing us to convert results on these parameters from either the RAR or $Q_2$ into a PPD of $\ag$ values. This is the distribution we would expect to see given the RAR or $Q_2$ inference, and may subsequently be compared to the posterior of a WBT such as that of~\citet{Banik_WBT} to assess the consistency of the MOND interpretation of these systems. For simplicity, following~\citeauthor{Banik_WBT} we fix $g_\text{e} = 2.142\times10^{-10}$ m s$^{-2}$ for this comparison, since varying this within the range allowed by \citet{klioner:2021} makes little difference to the results. Although straightforward, we do not compute \{$a_0$, shape\} posteriors from the inference of~\citeauthor{Banik_WBT}, as we do in Sec.~\ref{sec:results_q2} from the $Q_2$ measurement. This is in order not to restrict ourselves to those results, reflecting the less established nature of the WBT than the $Q_2$ measurement and the fact that other WBT analyses have reached substantially different conclusions~\citep{Hernandez, Hernandez_2, Hernandez_3, Hernandez_4, chae_wb, chae_wb_2}.

Like $Q_2$, $\ag$ goes to 0 in the limit shape $\rightarrow\infty$, so a flat prior on shape and the associated semi-infinite prior volume towards that limit would have made $\ag$ appear more strongly constrained by Cassini than it in fact is. This is cured by adopting a  flat prior on $Q_2$.
Note that the $\ag$ inference of \citet{Banik_WBT}, to which we compare the $\ag$ PPD from $Q_2$, used instead a uniform prior on $\ag$. We have checked that switching to this prior---or a log-uniform prior in $Q_2$ or $\ag$---makes little difference to our results.

\section{Results}
\label{sec:results}

\subsection{Constraints and predictions from the RAR}
\label{sec:results_rar}

\subsubsection{Fiducial mass-to-light ratio priors}


\begin{table*}
\begin{center}
\begin{tabular}{|c|c|c|c|c|c|c|c|c|c}
  \hline
  IF family & EFE model & shape & $a_0$ & $\e$ & $Q_2$ & $\sigma_{Q2}$ & $\ag$ & $\Delta$BIC & $\Delta$BIC(P)\\
  \hline
  \rule{0pt}{4ex}
  RAR IF & No EFE & --- & $1.03^{+0.03}_{-0.03}$ & ---  & $29.2^{+0.3}_{-0.4}$ & 8.7 & $0.74^{+0.03}_{-0.03}$ & 0 & 0\\
  \rule{0pt}{4ex}
  $\delta$ & No EFE & $0.97^{+0.04}_{-0.04}$ & $1.02^{+0.04}_{-0.04}$ & ---  & $29.4^{+0.4}_{-0.5}$ & 8.7 & $0.78^{+0.06}_{-0.06}$ & 11.1 & $-10.2$\\
  \rule{0pt}{4ex}
  $\delta$ & AQUAL global & $0.98^{+0.04}_{-0.04}$ & $1.03^{+0.04}_{-0.04}$ & $0.0017^{+0.001}_{-0.001}$  & $29.4^{+0.4}_{-0.5}$ & 8.7 & $0.77^{+0.07}_{-0.06}$ & 18.7 & 15.2\\
  \rule{0pt}{4ex}
  $\delta$ & AQUAL local & $1.14^{+0.05}_{-0.05}$ & $1.25^{+0.05}_{-0.05}$ & $\mathit{0.0048^{+0.0082}_{-0.0020}}$ & $30.2^{+0.5}_{-0.5}$ & 8.9 & $0.71^{+0.06}_{-0.06}$ & 1090 & 1480\\
  \rule{0pt}{4ex}
  $\delta$ & QUMOND global & $0.98^{+0.04}_{-0.04}$ & $1.03^{+0.04}_{-0.04}$ & $0.0049^{+0.0015}_{-0.0019}$  & $29.4^{+0.4}_{-0.5}$ & 8.7 & $0.77^{+0.07}_{-0.06}$ & 13.9 & $-1.98$\\
  \rule{0pt}{4ex}
  $\delta$ & QUMOND local & $1.08^{+0.05}_{-0.05}$ & $1.17^{+0.04}_{-0.04}$ & $\mathit{0.0050^{+0.0030}_{-0.0018}}$ & $29.9^{+0.5}_{-0.5}$ & 8.8 & $0.73^{+0.07}_{-0.06}$ & 1090 & 1470\\
  \rule{0pt}{4ex}
  $\gamma$ & No EFE & $1.03^{+0.07}_{-0.07}$ & $1.03^{+0.03}_{-0.03}$ & ---  & $29.1^{+0.4}_{-0.4}$ & 8.6 & $0.71^{+0.07}_{-0.07}$ & 5.93 & $-1.99$\\
  \rule{0pt}{4ex}
  $\gamma$ & AQUAL global & $1.04^{+0.07}_{-0.07}$ & $1.04^{+0.03}_{-0.03}$ & $0.0018^{+0.0010}_{-0.0010}$  & $29.2^{+0.4}_{-0.4}$ & 8.7 & $0.71^{+0.07}_{-0.07}$ & 14.7 & 14.9\\
  \rule{0pt}{4ex}
  $\gamma$ & AQUAL local & $1.14^{+0.06}_{-0.06}$ & $1.19^{+0.04}_{-0.04}$ & $\mathit{0.0048^{+0.0070}_{-0.0020}}$ & $30.8^{+0.4}_{-0.4}$ & 9.2 & $0.76^{+0.06}_{-0.06}$ & 1100 & 1510\\
  \rule{0pt}{4ex}
  $\gamma$ & QUMOND global & $1.04^{+0.07}_{-0.07}$ & $1.04^{+0.03}_{-0.03}$ & $0.0050^{+0.0015}_{-0.0018}$  & $29.2^{+0.4}_{-0.4}$ & 8.6 & $0.71^{+0.07}_{-0.06}$ & 9.69 & 8.32\\
  \rule{0pt}{4ex}
  $\gamma$ & QUMOND local & $1.11^{+0.07}_{-0.07}$ & $1.14^{+0.04}_{-0.04}$ & $\mathit{0.0050^{+0.0030}_{-0.0017}}$ & $30.2^{+0.4}_{-0.4}$ & 9.0 & $0.73^{+0.06}_{-0.06}$ & 1080 & 1490\\
  \rule{0pt}{4ex}
  $n$ & No EFE & $1.02^{+0.04}_{-0.04}$ & $1.08^{+0.04}_{-0.04}$ & ---  & $28.4^{+0.4}_{-0.4}$ & 8.4 & $0.87^{+0.07}_{-0.06}$ & 17.7 & 23.3\\
  \rule{0pt}{4ex}
  $n$ & AQUAL global & $1.03^{+0.04}_{-0.04}$ & $1.09^{+0.04}_{-0.04}$ & $0.0018^{+0.0009}_{-0.0010}$ & $28.4^{+0.4}_{-0.4}$ & 8.4 & $0.86^{+0.07}_{-0.06}$ & 26.3 & 33.2\\
  \rule{0pt}{4ex}
  $n$ & AQUAL local & $1.19^{+0.06}_{-0.04}$ & $1.31^{+0.05}_{-0.05}$ & $\mathit{0.0048^{+0.0094}_{-0.0020}}$ & $29.4^{+0.5}_{-0.5}$ & 8.7 & $0.79^{+0.07}_{-0.06}$ & 1100 & 1490\\
  \rule{0pt}{4ex}
  $n$ & QUMOND global & $1.03^{+0.04}_{-0.04}$ & $1.09^{+0.04}_{-0.04}$ & $0.0049^{+0.0014}_{-0.0017}$  & $28.4^{+0.4}_{-0.4}$ & 8.4 & $0.86^{+0.07}_{-0.06}$ &  20.9 & 25.6\\
  \rule{0pt}{4ex}
  $n$ & QUMOND local & $1.12^{+0.05}_{-0.05}$ & $1.23^{+0.04}_{-0.04}$ & $\mathit{0.0051^{+0.0032}_{-0.0018}}$ & $29.1^{+0.4}_{-0.5}$ & 8.6 & $0.82^{+0.07}_{-0.06}$ & 1080 & 1470\\
  \hline
\end{tabular}
\caption{Constraints on RAR (or, equivalently, algebraic MOND) parameters and goodness-of-fit statistics for the fiducial SPARC $M/L$ model. The parameter $a_0$ has units of $10^{-10}\:\mathrm{m}\,\mathrm{s}^{-2}$ and $Q_2$ of $10^{-27}\:\mathrm{s}^{-2}$. We show all combinations of IF family (Eq.~\ref{eq:if_fams}) and EFE model (Eqs.~\ref{eq:efe_fo} and \ref{eq:efe_an}).
For the models with galaxy-specific (``local'') EFE the quoted $\e$ constraints---written in italics---describe the median stacked posteriors over all galaxies, implicitly including the maximum-clustering prior. The $Q_2$ (Eq.~\ref{eq:Q2}) and $\ag$ (Eq.~\ref{eq:ag}) columns summarise the posterior predictive distributions of the quadrupole and WB test statistic from the RAR fits, while $\sigma_{Q2}$ is the number of sigma tension between the $Q_2$ prediction and the Cassini measurement. For reference, the first row shows the RAR IF fit without the EFE; the BIC values (defined using either the maximum-likelihood or maximum-posterior points) are shown relative to this.
}
\label{tab:results_1}
\end{center}
\end{table*}

\begin{table*}
\begin{center}
\begin{tabular}{|c|c|c|c|c|c|c|c|c|c|c}
  \hline
  $M/L$ model & EFE model & shape ($\delta$) & $a_0$ & $\mathbf{\sig}$ & $\mu_\text{d}$ & $\mu_\text{b}$ & $Q_2$ & $\sigma_{Q2}$ & $\ag$ & $\Delta$BIC\\
  \hline
  \rule{0pt}{4ex}
  Free hyper & No EFE & $1.28^{+0.06}_{-0.06}$ & $1.04^{+0.03}_{-0.03}$ & $0.034$ & $0.71^{+0.03}_{-0.03}$ & $0.62^{+0.04}_{-0.03}$ & $25.5^{+0.8}_{-0.9}$ & 7.2 & $0.40^{+0.05}_{-0.05}$ & $-53.0$\\
  \rule{0pt}{4ex}
  Free hyper & AQUAL local & $1.50^{+0.08}_{-0.07}$ & $1.21^{+0.04}_{-0.04}$ & $0.032$ & $0.72^{+0.03}_{-0.03}$ & $0.62^{+0.04}_{-0.03}$ & $25.6^{+0.9}_{-0.9}$ & 7.2 & $0.34^{+0.05}_{-0.05}$ & 1000\\
  \rule{0pt}{4ex}
  Const hyper & No EFE & $1.24^{+0.06}_{-0.06}$ & $1.05^{+0.03}_{-0.03}$ & $0.034$ & $0.68^{+0.02}_{-0.02}$ & $0.69^{+0.02}_{-0.02}$ & $26.3^{+0.7}_{-0.8}$ & 7.5 & $0.44^{+0.05}_{-0.05}$ & $-46.9$\\
  \rule{0pt}{4ex}
  Const hyper & AQUAL local & $1.45^{+0.07}_{-0.07}$ & $1.23^{+0.04}_{-0.04}$ & $0.032$ & $0.69^{+0.02}_{-0.02}$ & $0.70^{+0.02}_{-0.02}$ & $26.6^{+0.8}_{-0.9}$ & 7.6 & $0.39^{+0.05}_{-0.05}$ & 1000\\
  \rule{0pt}{4ex}
  Free unif & No EFE & $1.28^{+0.06}_{-0.06}$ & $1.10^{+0.03}_{-0.03}$ & $0.031$ & $\mathit{0.73^{+0.72}_{-0.40}}$ & $\mathit{0.68^{+0.41}_{-0.27}}$ & $26.5^{+0.7}_{-0.8}$ & 7.6 & $0.44^{+0.05}_{-0.05}$ & $-440$\\
  \rule{0pt}{4ex}
  Free unif & AQUAL local & $1.52^{+0.08}_{-0.07}$ & $1.24^{+0.04}_{-0.04}$ & $0.029$ & $\mathit{0.74^{+0.70}_{-0.38}}$ & $\mathit{0.60^{+0.53}_{-0.19}}$ & $25.9^{+0.9}_{-0.9}$ & 7.3 & $0.34^{+0.05}_{-0.05}$ & 637\\
  \rule{0pt}{4ex}
  Const unif & No EFE & $1.06^{+0.04}_{-0.04}$ & $1.18^{+0.04}_{-0.04}$ & $0.032$ & $\mathit{0.58^{+0.73}_{-0.33}}$ & $\mathit{0.73^{+0.58}_{-0.27}}$ & $30.1^{+0.4}_{-0.5}$ & 8.9 & $0.75^{+0.06}_{-0.06}$ & $-274$\\
  \rule{0pt}{4ex}
  Const unif & AQUAL local & $1.24^{+0.06}_{-0.05}$ & $1.39^{+0.05}_{-0.05}$ & $0.031$ & $\mathit{0.57^{+0.72}_{-0.29}}$ & $\mathit{0.74^{+0.63}_{-0.28}}$ & $30.8^{+0.5}_{-0.6}$ & 9.1 & $0.68^{+0.06}_{-0.06}$ & 823\\
  \rule{0pt}{4ex}
  No bulge & No EFE & $1.97^{+0.22}_{-0.18}$ & $0.99^{+0.03}_{-0.03}$ & $0.041$ & $0.81^{+0.03}_{-0.03}$ & --- & $14.3^{+3.0}_{-2.9}$ & 2.7 & $0.05^{+0.04}_{-0.04}$ & 2140\\
  \rule{0pt}{4ex}
  No bulge & AQUAL local & $2.49^{+0.33}_{-0.27}$ & $1.10^{+0.04}_{-0.04}$ & $0.039$ & $0.81^{+0.03}_{-0.03}$ & --- & $10.9^{+3.2}_{-2.9}$ & 1.9 & $0.00^{+0.03}_{-0.02}$ & 2940\\
  \rule{0pt}{4ex}
  Only bulge & No EFE & $0.99^{+0.06}_{-0.06}$ & $1.30^{+0.08}_{-0.07}$ & $0.026$ & $0.59^{+0.05}_{-0.05}$ & $0.68^{+0.04}_{-0.04}$ & $31.7^{+0.5}_{-0.6}$ & 9.4 & $0.99^{+0.13}_{-0.12}$ & $-191$\\
  \rule{0pt}{4ex}
  Only bulge & AQUAL local & $1.19^{+0.08}_{-0.07}$ & $1.57^{+0.10}_{-0.09}$ & $0.024$ & $0.62^{+0.05}_{-0.04}$ & $0.69^{+0.04}_{-0.04}$ & $32.7^{+0.7}_{-0.8}$ & 9.6 & $0.87^{+0.11}_{-0.10}$ & $-105$\\
  \hline
\end{tabular}
\caption{As Table~\ref{tab:results_1}, but focusing on the $\delta$-family and models either without an EFE or with a galaxy-specific AQUAL EFE with maximum-clustering prior on $\eN$, allowing the $M/L$ model to vary. The models considered are: 1) unconstrained Gaussian hyperpriors on $\Ud$ and $\Ub$, 2) as (1) but requiring the centre of the $\Ub$ prior to exceed that of the $\Ud$ prior, 3) free uniform priors on $\Ud$ and $\Ub$, 4) as (3) but requiring $\Ub>\Ud$ galaxy-by-galaxy, 5) as (2) but excluding galaxies with bulges, and 6) as (2) but excluding galaxies without bulges. For models with Gaussian hyperpriors on $\Ud$ and $\Ub$, the $\mu_\text{d}$ and $\mu_\text{b}$ columns show the mean and 68\% confidence intervals of their centres; for the other models (italicised) they instead show the mean and 68\% confidence intervals of the mean $\Ud$ and $\Ub$ values across the posteriors of all galaxies. The uncertainty on $\sig$ (measured in dex) is $\pm 0.001$ in all cases. $\Delta$BIC is again defined relative to the top row of Table~\ref{tab:results_1}.
}
\label{tab:results_2}
\end{center}
\end{table*}

\begin{table*}
\begin{center}
\begin{tabular}{|c|c|c|c|c|c|c|c|c|c}
  \hline
  $M/L$ model & shape ($n$) & $a_0$ & $\mathbf{\sig}$ & $\mu_\text{d}$ & $\mu_\text{b}$ & $Q_2$ & $\sigma_{Q2}$ & $\ag$ & $\Delta$BIC\\
  \hline
  \rule{0pt}{4ex}
  Fiducial & $0.78^{+0.03}_{-0.03}$ & $0.96^{+0.05}_{-0.05}$ & $0.046$ & 0.5 & 0.7 & $29.2^{+0.3}_{-0.2}$ & 8.7 & $1.34^{+0.09}_{-0.08}$ & 2050\\
  \rule{0pt}{4ex}
  Free hyper & $1.57^{+0.09}_{-0.09}$ & $1.12^{+0.03}_{-0.03}$ & $0.042$ & $0.97^{+0.03}_{-0.03}$ & $0.64^{+0.04}_{-0.04}$ & $22.8^{+1.0}_{-1.0}$ & 6.3 & $0.33^{+0.05}_{-0.05}$ & 2290\\
  \rule{0pt}{4ex}
  Free unif & $1.34^{+0.08}_{-0.07}$ & $1.10^{+0.04}_{-0.03}$ & $0.041$ & $\mathit{0.88^{+0.76}_{-0.40}}$ & $\mathit{0.64^{+0.43}_{-0.25}}$ & $25.2^{+0.8}_{-0.9}$ & 7.1 & $0.49^{+0.06}_{-0.06}$ & 2120\\
  \rule{0pt}{4ex}
  No bulge & $1.93^{+0.19}_{-0.17}$ & $1.01^{+0.03}_{-0.03}$ & $0.049$ & $1.08^{+0.04}_{-0.04}$ & --- & $17.5^{+2.4}_{-2.4}$ & 3.8 & $0.15^{+0.05}_{-0.04}$ & 3680\\
  \hline
\end{tabular}
\caption{As Table~\ref{tab:results_2}, but using the modified gravity prescription of Eq.~\ref{eq:bm} for calculating the MONDian radial acceleration $g$ rather than the algebraic MOND relation. Here we use the $n$-family without EFE in all cases.}
\label{tab:results_MG}
\end{center}
\end{table*}

We begin with the fiducial SPARC $M/L$ model \citep{Schombert}. Table~\ref{tab:results_1} shows in this case the constraints on the RAR parameters for each IF family and EFE model. We do not show $\sig$, which is between 0.033 and 0.035 dex for all models with an uncertainty of 0.001. For the local EFE models where $\e$ varies per galaxy, we show the median and 68\% confidence intervals of the median $\e$ values across the posteriors of all galaxies; these entries are italicised to distinguish them from the qualitatively different global $\e$ constraints appearing in the same column. For each model we also summarise the PPDs of $Q_2$ and $\ag$ by their median and $68$ per cent confidence interval, including the tension with the Cassini measurement in the case of $Q_2$. This is calculated assuming a Gaussian distribution of the lower uncertainty.

The final two columns describe the goodness-of-fit, specifically the the Bayesian information criterion (BIC) relative to the reference RAR IF shown in the top row. The BIC is the limit of the Bayesian evidence (the probability of the data given the model) when the posterior is modelled as a Gaussian around the maximum \textit{a posteriori} point and the number of data points greatly exceeds the number of free parameters. Although neither of these assumptions are manifestly true in our case, the BIC still functions as a useful model comparison heuristic by trading off the accuracy of a model with its complexity in terms of number of free parameters. It is given by \citep{BIC}
\begin{equation}
\text{BIC} \equiv k\ln(N) - 2\ln(\hat{\mathcal{L}}),
\end{equation}
where $k$ is the number of free parameters, $N$ the number of data points and $\hat{\mathcal{L}}$ the maximum likelihood value. The evidence is proportional to $\exp(-\text{BIC}/2)$. On the Jeffreys scale~\citep{JS}, an evidence ratio in excess of 100 ($|\Delta\text{BIC}|>9.21$) indicates ``decisive'' evidence in favour of the higher-evidence, lower-BIC model.

In the limit of much data, the likelihood approximates the posterior because the prior does not scale with the number of datapoints while the likelihood does. In our case, however, a better estimator of model quality may be achieved by replacing the maximum-likelihood with maximum-posterior value. This leads us to define
\begin{equation}
\text{BIC(P)} \equiv k\ln(N) - 2\ln(\hat{\mathcal{P}}),
\end{equation}
for maximum posterior value $\hat{\mathcal{P}}$, which we also show. Note that since all log-probability values are negative, introducing additional parameters must increase BIC(P) relative to BIC. For models with the same priors, however, BIC(P) may provide a better model comparison statistic because it is the relative posterior probabilities of the models that are important, not only the likelihoods they assign the data.

\begin{figure*}
  \centering
  \includegraphics[width=0.95\textwidth]{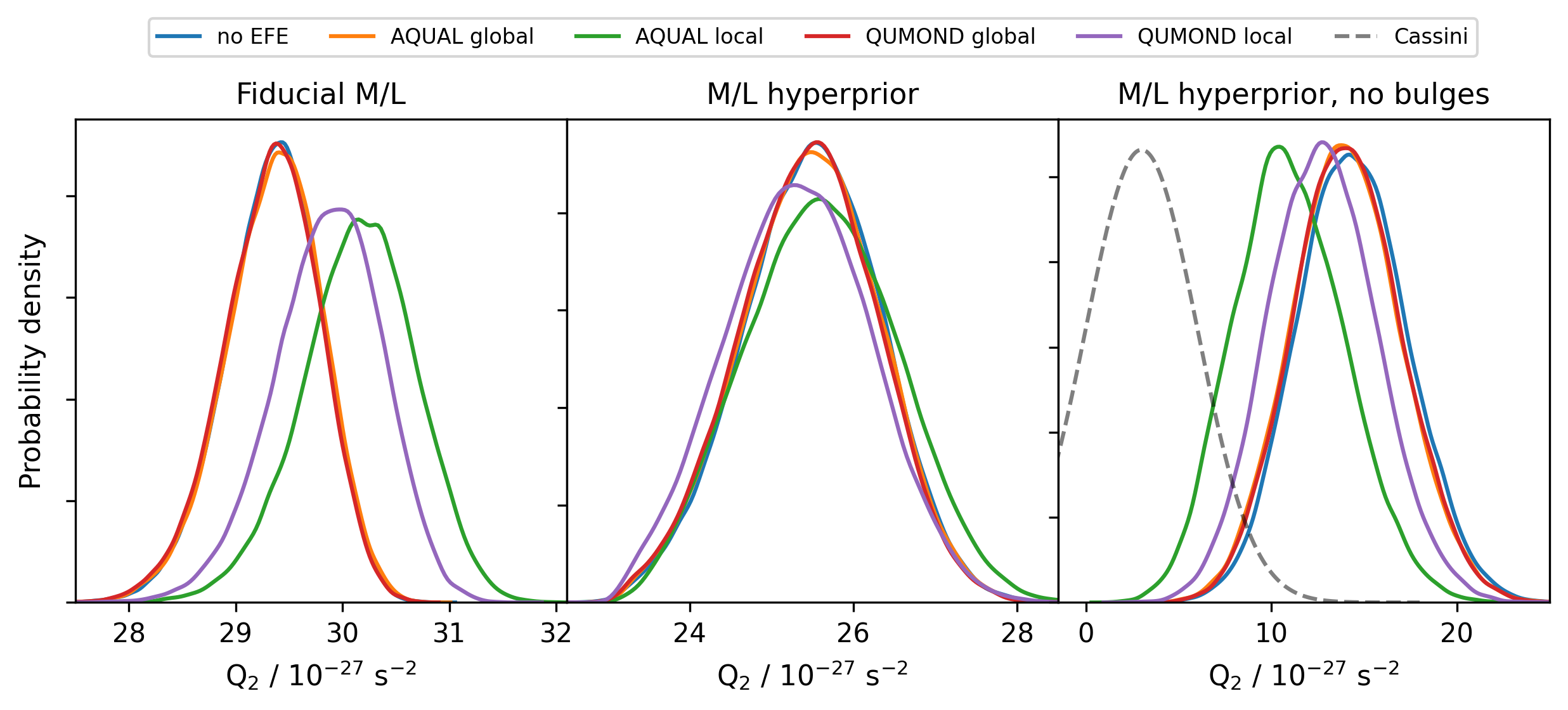}
  \caption{PPD of $Q_2$ from the RAR for the $\delta$-family under the five different EFE models. \emph{Left:} Fiducial SPARC $M/L$ model; \emph{centre:} Gaussian hyperprior model; \emph{right:} as centre but excluding galaxies with bulges. The ``local'' EFE models, with a separate $\e$ per galaxy, use the maximum-clustering prior. The constraint on $Q_2$ from Cassini is visible in dashed grey in the right panel.}
  \label{fig:q2_ppd}
\end{figure*}

\begin{figure*}
  \centering
  \includegraphics[width=0.95\textwidth]{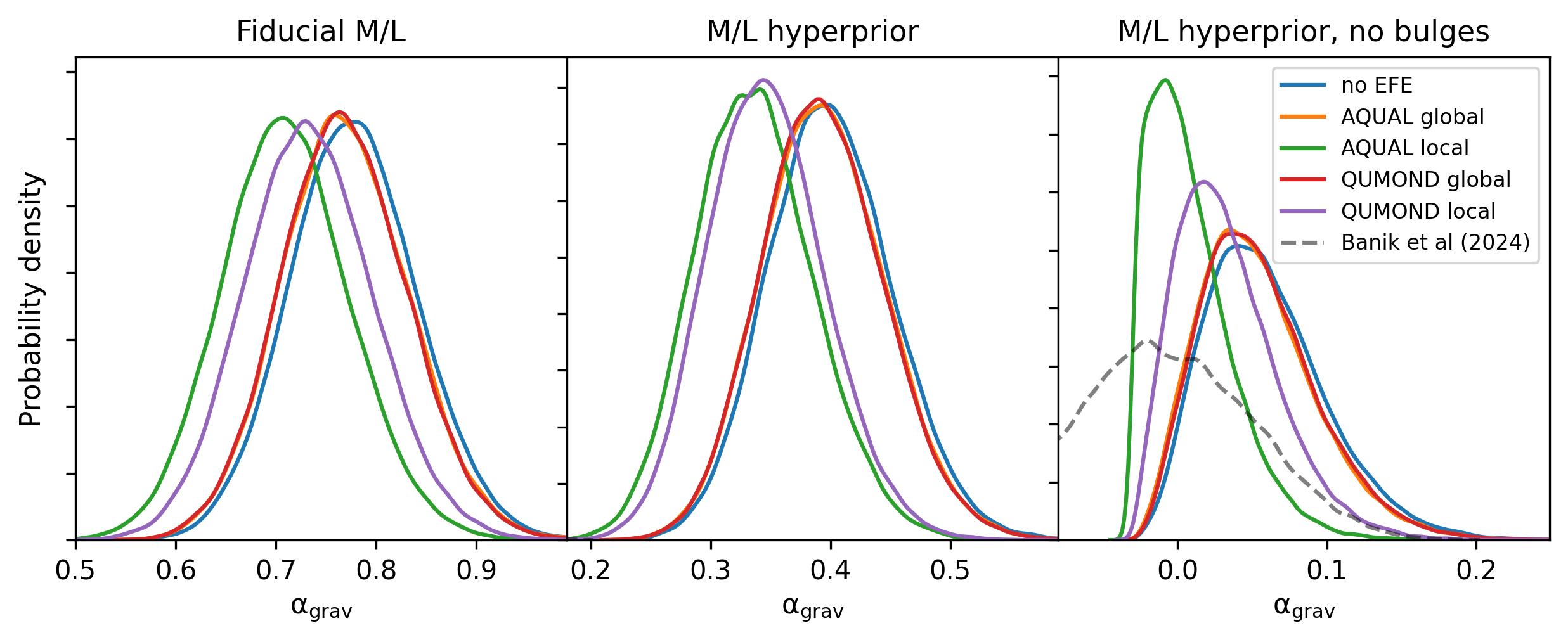}
  \caption{As Fig.~\ref{fig:q2_ppd}, but for $\ag$ rather than $Q_2$. The constraint on $\ag$ from the WBT of~\citet{Banik_WBT} is visible in dashed grey in the right panel.}
  \label{fig:agrav_rar}
\end{figure*}

We highlight three results from this investigation:
\begin{itemize}
\item $a_0$ and shape are in all cases well constrained and largely consistent between the models. They show the RAR to have a transition location consistent with recent studies (e.g. table~3 of \citealt{uRAR})
and a transition sharpness closely approximating the Simple (shape=1 in the $n$-family) and RAR (shape=1 in the $\delta$- and $\gamma$-families) IFs.  There is therefore at most weak evidence for a more general IF relative to Simple or RAR. A corollary is that the family with which one extends these IFs is not particularly important, allowing us to focus our remaining analyses on a single one. We choose the $\delta$-family because it is the simplest that includes the most popular RAR function. Some historical context to our $a_0$ constraints may be found in Sec.~\ref{sec:conc}.
\item Allowing local gravitational field strengths in the EFE is strongly disfavoured by the BIC relative to the no-EFE model due to the addition of 147 free parameters, and allowing a single global field strength is mildly disfavoured. Our analysis does not therefore show strong evidence for the existence of the EFE in the SPARC RAR.
This can also be seen from the fact that the $\eN$ values of the global-EFE models are all consistent with 0 within 2$\sigma$. It is however curious to note that while the AQUAL constraints on a global $\e$ are significantly below the expectation from the maximum-clustering prior (and hence also the galaxy-by-galaxy $\e$ constraints from the local EFE model), the QUMOND constraints are almost identical. This suggests that the RC constraints accord better with the large-scale structure priors in QUMOND than AQUAL, which is supported by the lower BIC and BIC(P) values for the QUMOND EFE models. These results differ from those of~\citet{Chae_1,Chae_2,Chae_3} due to their focus on specific galaxies and ours on the overall RAR, and our different fitting  and goodness-of-fit assessment procedures.
\item $Q_2$ and $\ag$ are clearly non-zero in all cases, indicating that deviation from Newtonian gravity in the SS quadrupole and WB dynamics should be detected if these models are correct. Note however that the expected $\ag$ is less than unity---the result for the Simple IF at $a_0=1.2\times10^{-10}$ m s$^{-2}$---in all cases, largely due to the lower preferred $a_0$ value.
The tensions with the quadrupole measurement of Eq.~\ref{eq:Q2_measurement} are strong, $\sim$8--9$\sigma$. The left panels of Figs.~\ref{fig:q2_ppd} and~\ref{fig:agrav_rar} show the full PPDs of $Q_2$ and $\ag$ respectively.
\end{itemize}

\subsubsection{Extended mass-to-light ratio priors}
\label{sec:ML_ext}

\begin{figure*}
  \centering
  \includegraphics[width=0.45\textwidth]{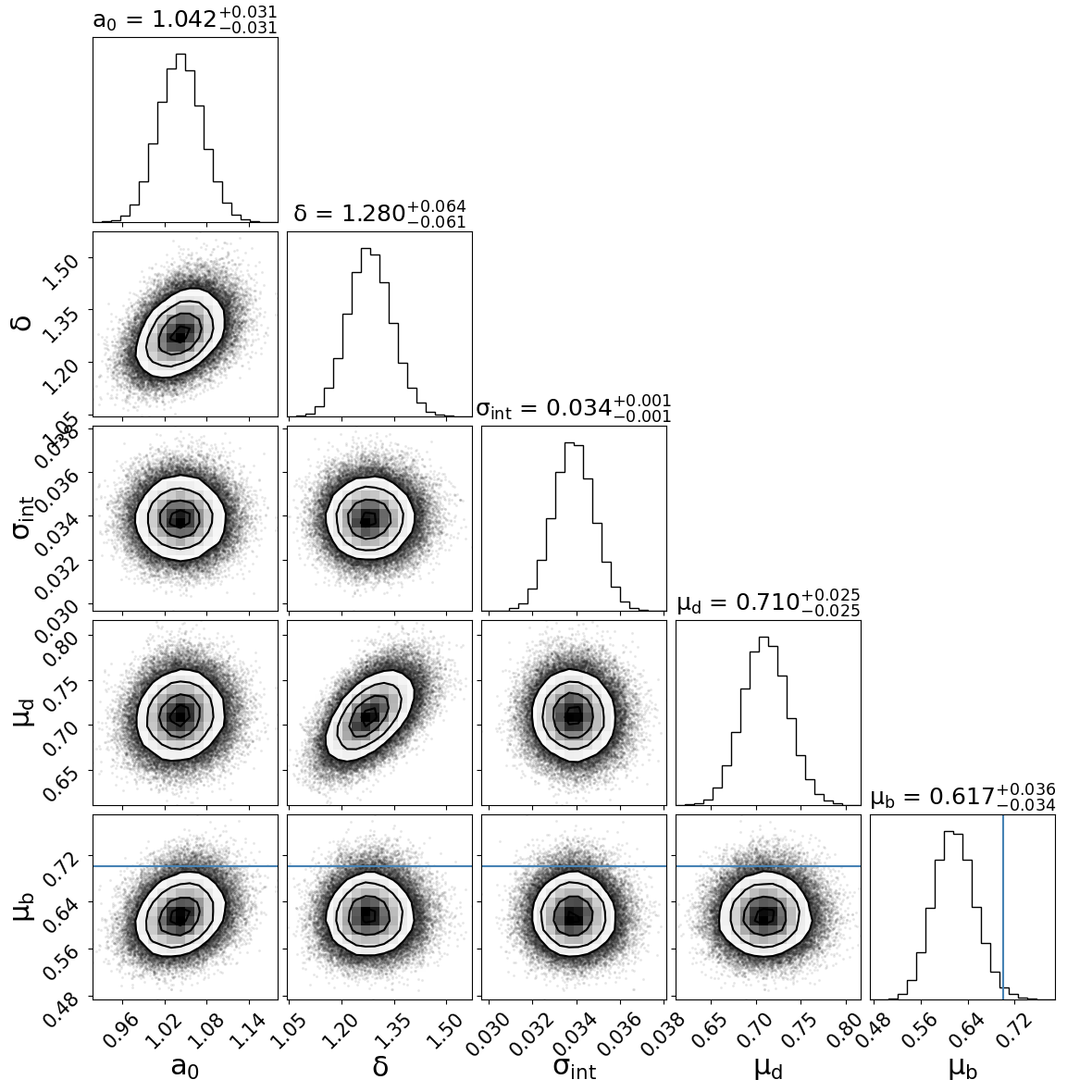}
  \includegraphics[width=0.45\textwidth]{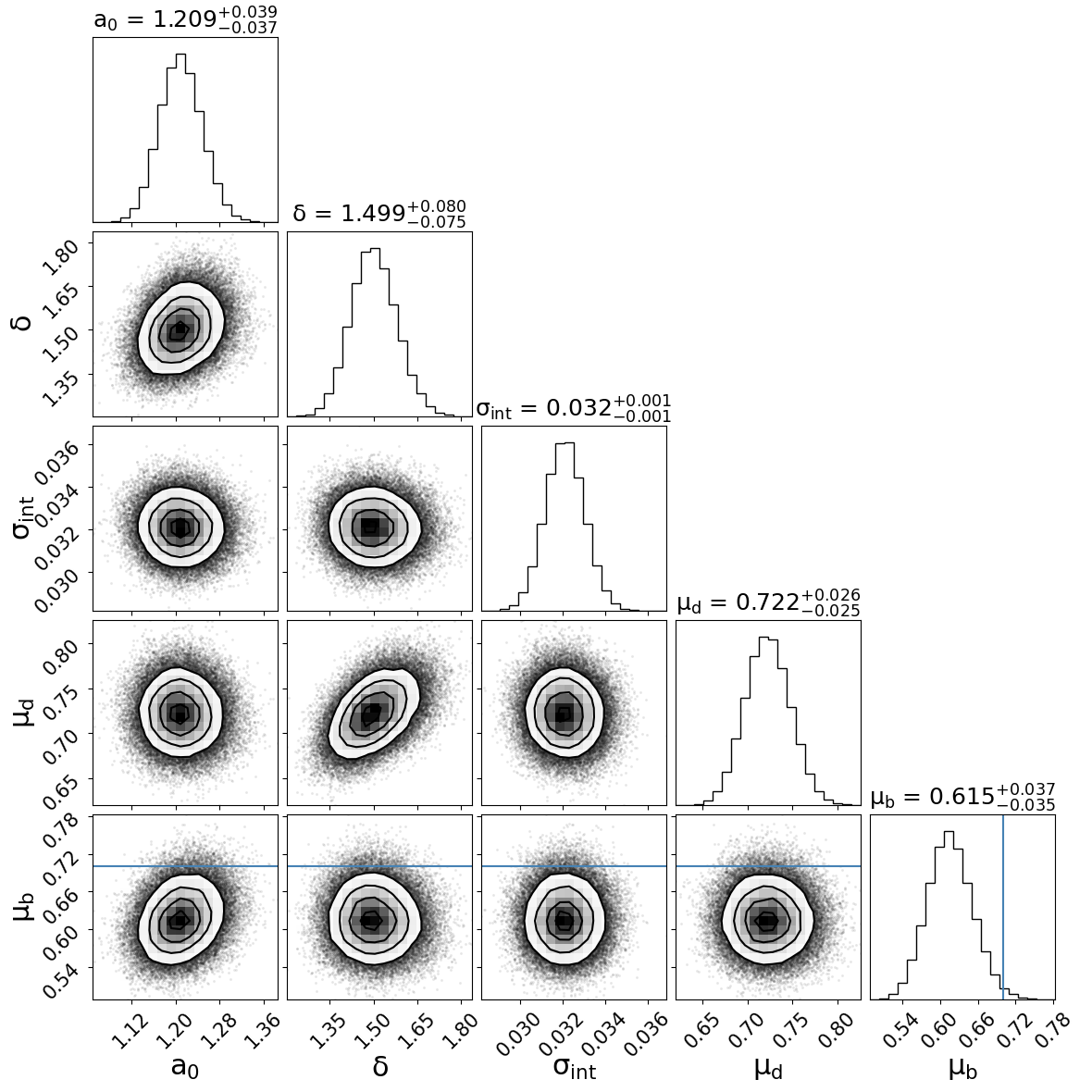}
  \caption{Partial corner plots of the RAR inference using the $\delta$-family of IFs and the Gaussian hyperprior model for $\Ud$ and $\Ub$, for the case of no EFE (\emph{left}) and AQUAL with maximum-clustering $\e$ prior (\emph{right}). The truth line shows $\mu_\text{b}=0.7$, the fiducial SPARC value (the corresponding $\mu_\text{d}=0.5$ is off the plot).}
  \label{fig:corner}
\end{figure*}

We begin our investigation of more general $M/L$ models with a Gaussian hyperprior model in which the centres of the lognormal priors on $\Ud$ and $\Ub$ are free parameters to be inferred alongside $a_0$ and shape. These results comprise the first two rows of Table~\ref{tab:results_2} (for the $\delta$-family). We find the best-fit values of the hyperprior centres to be $\mu_\text{d}\approx0.7$ and $\mu_\text{b}\approx0.6$ regardless of the IF family and EFE model used. This provides a significantly better fit than the fiducial SPARC model, and the BIC shows that the addition of these two extra parameters is clearly warranted by the data. These values are however highly unexpected from a stellar population modelling point of view: in dwarfs one should if anything expect $\Ud<0.5$ (F. Lelli, priv. comm.), while the older stars in the bulge should in all cases have higher $M/L$ than those in the disk. In Fig.~\ref{fig:corner} we show partial corner plots of the Gaussian hyperprior model for the $\delta$-family and two different EFE models: no EFE (left panel) and AQUAL EFE with maximum-clustering prior (right panel).

In addition to improving the goodness of fit, allowing this extra freedom in $\Ud$ and $\Ub$ significantly increases $\delta$ from $\sim$1 to up to $\sim$1.$3-1$.5, corresponding to a sharper deep-MOND-to-Newtonian transition. As a result, the predicted $Q_2$ and $\ag$ values are significantly reduced, as shown in the central panels of Figs.~\ref{fig:q2_ppd} and~\ref{fig:agrav_rar}. The effect is most pronounced using the AQUAL local EFE model. However, although the tensions with Cassini and the WBT of~\citet{Banik_WBT} are eased relative to the fiducial $M/L$ model they remain significant, at the $\sim$7$\sigma$ level for $Q_2$.

The other $M/L$ models of Sec.~\ref{sec:rar_inference}---shown in the remaining rows of Table~\ref{tab:results_2}---tell similar stories. Allowing $\Ud$ and $\Ub$ to float freely yields $\Ud>\Ub$ with a corresponding significant improvement in goodness-of-fit. The fact that BIC is significantly lower for the free uniform than free hyperprior model, along with the best-fit scatter on $\Ud$ and $\Ub$ between galaxies being considerably larger than 25 per cent (italicised entries), suggests that the $M/L$ values may not in fact possess the degree of similarity expected in the SPARC error model, at least under the assumption of a fixed RAR. Forcing $\mu_\text{d}<\mu_\text{b}$ or $\Ud<\Ub$ results in $\mu_\text{d} \approx \mu_\text{b}$ and $\Ud \approx \Ub$ respectively, with worsened goodness-of-fit and increased $Q_2$ and $\ag$. We omit the $\e$ constraints in this table because they are very similar to those in Table~\ref{tab:results_1}, and show only $\Delta$BIC (not $\Delta$BIC(P)) for goodness-of-fit.

\begin{figure}
  \centering
  \includegraphics[width=0.495\textwidth]{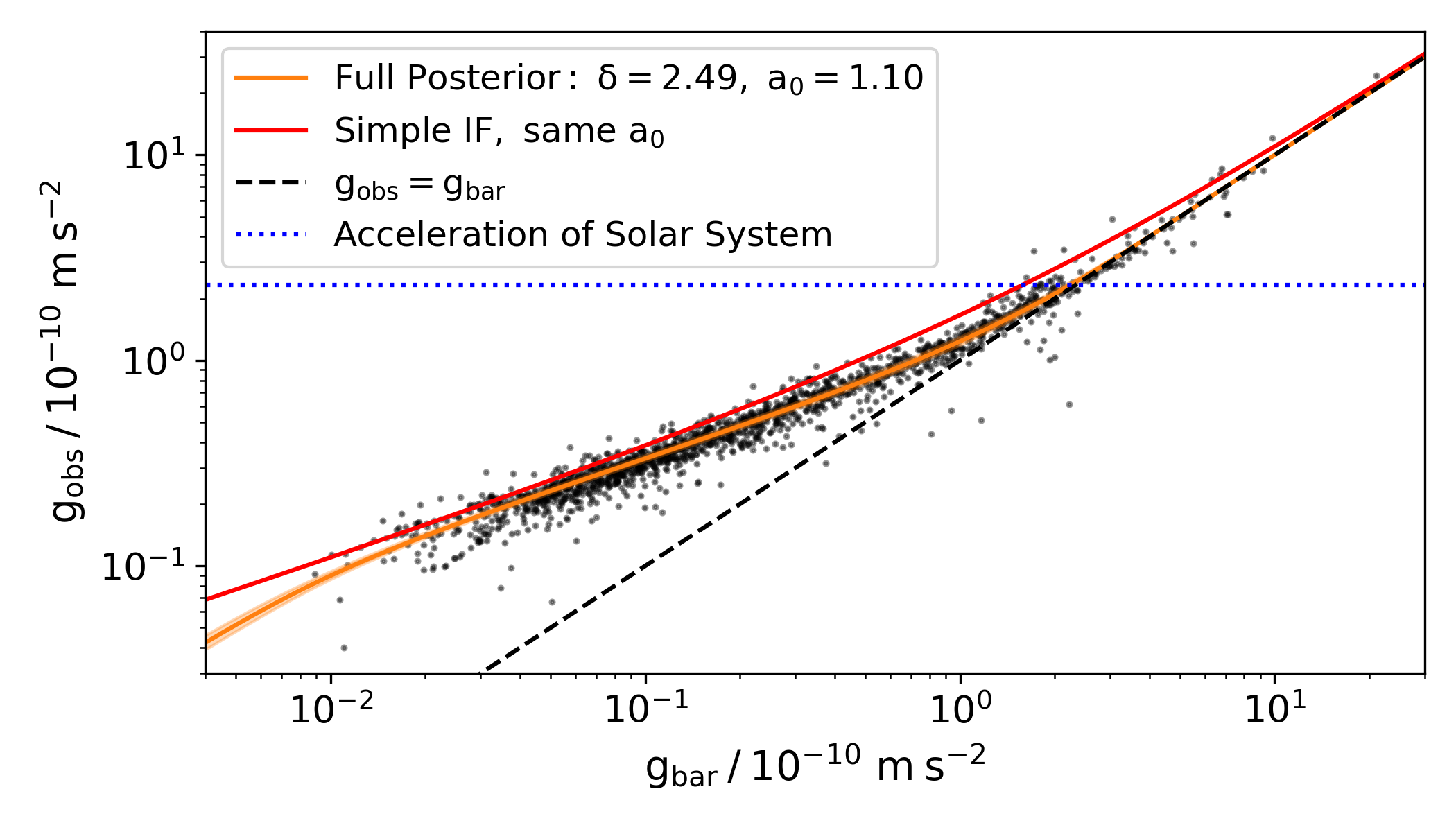}
  \caption{SPARC data points transformed according to the best-fit parameters for the $\delta$-family, local AQUAL EFE and $M/L$ hyperprior model removing galaxies with bulges. The orange line and band show the best-fit RAR and its 95 per cent confidence limit (producing a very small boost to gravity in the SS), while the red line shows the Simple IF with the same $a_0$.}
  \label{fig:big_shape}
\end{figure}

Even more drastic changes are apparent in the final four rows of Table~\ref{tab:results_2}, where we remove either galaxies with bulges or galaxies without bulges (in both cases using a Gaussian hyperprior for the remaining $M/L$s). Removing the galaxies with bulges causes an immense increase in best-fit shape---reaching $\delta=2.5$ for the AQUAL local EFE model with maximum-clustering prior---and a more modest decrease in $a_0$. The RAR points transformed according to the best-fit galaxy parameter values with the best-fit $\delta$-family fit overplotted is shown in Fig.~\ref{fig:big_shape}, illustrating the very sharp transition between the Newtonian and deep-MOND regimes.
The SS, with acceleration $2.32 \times10^{-10}$ m s$^{-2}$ shown by the horizontal blue line, is almost fully Newtonian. (In contrast, the Simple IF with the same $a_0$, shown in red, gives a large MOND boost at that acceleration.) This reduces the expected $Q_2$ almost to zero as shown in the right panel of Fig.~\ref{fig:q2_ppd}, bringing it approximately into consistency with the Cassini measurement.
The expected $\ag$ is concomitantly reduced to 0 (right panel of Fig.~\ref{fig:agrav_rar}), making it fully consistent with the null result of~\citet{Banik_WBT}. The overall goodness-of-fit is however reduced, as shown by larger $\Delta$BIC and $\Delta$BIC(P) values.
We note that the use of a Gaussian hyperprior for the mean of the disc mass-to-light ratio $\mu_\text{d}$ is important for the great increase in the best-fit shape parameter when removing galaxies with bulges; there is almost no such increase if the fiducial SPARC $M/L$ model is used instead, with best-fit $\delta$ returning to $\sim$1.
Conversely, removing galaxies without bulges lowers shape, increases $a_0$,
brings back a strong tension with the $Q_2$ measurement, and predicts $\ag\approx1$.

\begin{figure}
  \centering
  \includegraphics[width=0.45\textwidth]{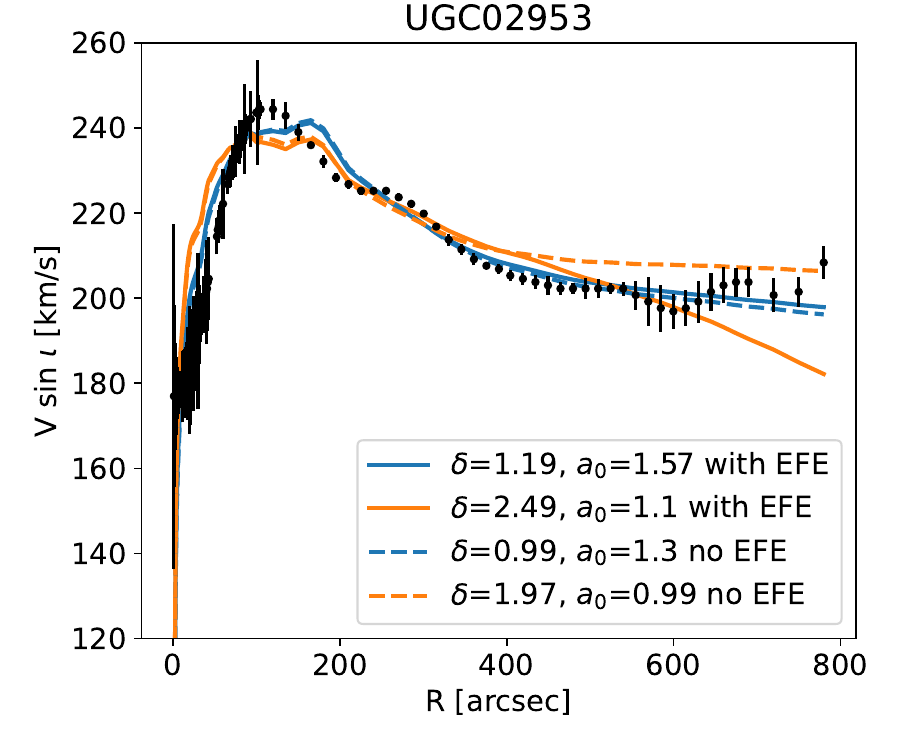}
  \caption{Example of fits for an individual bulgey galaxy, UGC~2953. The blue curves correspond to a fit where $a_0$ and shape are fixed to their best-fit values based on the RAR of bulgey galaxies only (last lines of Table~\ref{tab:results_2}). The orange curves correspond instead to a fit where $a_0$ and shape are fixed to their best-fit values based on the RAR of bulgeless galaxies only. The solid lines correspond to fits including the AQUAL-local EFE, while the dashed lines correspond to fits without EFE. The fitted parameters (distances, inclination and $\Upsilon$s) are additionally pushed far from their priors for the orange fits. This illustrates that bulgey galaxies are not well described by the RAR obtained from bulge-free galaxies alone, which could otherwise satisfy the $Q_2$ and WBT constraints.}
  \label{fig:comp_fit}
\end{figure}

\begin{figure}
  \centering
  \includegraphics[width=0.45\textwidth]{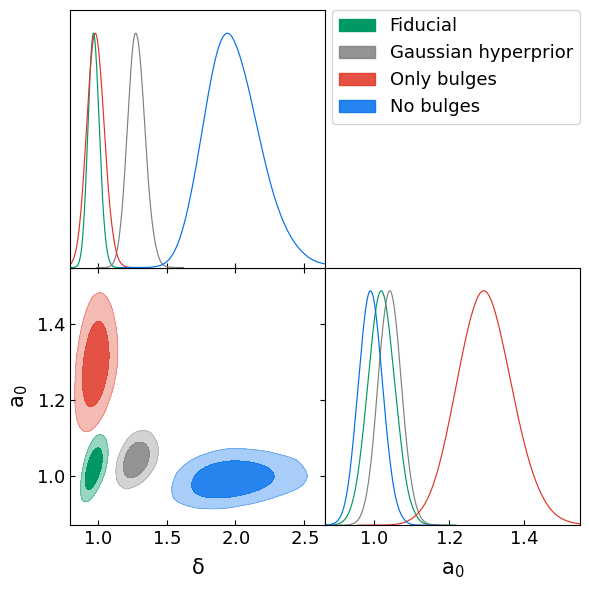}
  \caption{Comparison of constraints on $a_0$ and $\delta$ from the RAR under various $M/L$ models, assuming no EFE. In the context of MOND, the discordance between the ellipses of the fiducial, Gaussian hyperprior, no-bulge and bulge-only fits seem to hint at systematics in the SPARC data, subtleties related to the underlying MOND theory and/or significant issues with the fiducial $M/L$ priors. This disagreement cannot be resolved by changing the EFE model.}
  \label{fig:getdist_bulcomp}
\end{figure}

These results indicate an incompatibility between the separate RAR fits of bulgey and bulge-free galaxies. Fig.~\ref{fig:comp_fit} illustrates this by showing various fits for one bulgey galaxy, UGC~2953. The fits obtained when fixing the values of $a_0$ and shape parameter to their best fit values obtained from bulge-free galaxies ($\delta=1.97$, $a_0 = 0.99 \times 10^{-10} \, \mathrm{m}\,\mathrm{s}^{-2}$ without EFE, and $\delta=2.49$, $a_0 = 1.1 \times 10^{-10} \, \mathrm{m}\,\mathrm{s}^{-2}$ with EFE) are quite poor. In addition, the best-fit distance, inclination and $\Upsilon$ values are far from their priors. It therefore appears unlikely that the high shape values preferred by the bulge-free galaxies can explain the dynamics of bulgey galaxies.
To provide another angle on this discrepancy, we show in Fig.~\ref{fig:getdist_bulcomp} the \{$a_0$, shape\} constraints without EFE from the fiducial SPARC model, the free hyperprior model and the models with bulgey or bulge-free galaxies removed (using a Gaussian hyperprior for the remaining $M/L$s). Each model is in clear tension with the others.
This could be due either to the modelling of the bulge and disc light within the SPARC data reduction pipeline, or to inaccuracy of the RAR modelling of these components.
For example, the violation of axisymmetry caused by a significant bulge or bar---or the dependence on past trajectories and exact orbital structure in modified inertia formulations---may greatly modify the effective MOND force law. Using the fiducial $M/L$ model rather than the Gaussian hyperprior for the no-bulge and bulge-only cases the best-fit $\delta$ values are reduced, to $\sim$1 for the former and $\sim$0.9 for the latter, but they remain clearly discrepant in $a_0$. We leave further investigation of this issue to further work.


\subsubsection{AQUAL modified gravity RAR fit}
\label{sec:results_MG}

Our fiducial analysis above uses the algebraic MOND relation, Eq.~\ref{eq:g}, which is also the usual form of the RAR. While this is an exact prediction in modified inertia formulations of the theory for circular orbits, modified gravity formulations necessarily deviate from it to a (small) extent that depends on the geometry of the system \citep[see][]{FamaeyMcG}. As we are interested here in testing modified gravity theories (specifically AQUAL and QUMOND) in which the IF is the same between galaxy RCs, the Solar System, and nearby wide binaries, it is important to check that our results are not significantly affected by this deviation from the algebraic relation. To this end, we employ here the AQUAL approximation found by \citet[][eq 25]{Brada_Milgrom} for flat disks:
\begin{equation}\label{eq:bm}
g = \frac{g_N}{\mu\left(\frac{g_N^+}{a_0} \nu\left(\frac{g_N^+}{a_0}\right)\right)}
\end{equation}
where $g_N^+ \equiv \left(g_N^2 + (2\pi G\Sigma)^2\right)^{1/2}$, for a total baryonic surface density $\Sigma$ at each point in the disk. Here $\mu$ is the usual alternative form of the IF already described in Sect.~\ref{sec:WBT_method}, defined by $\nu(x\mu(x)) \equiv 1/\mu(x)$. Eq.~\ref{eq:bm} is appropriate for the radial acceleration predicted by AQUAL within razor-thin axisymmetric systems; we caution that its accuracy will be lower for galaxies with bulges.

The baryonic surface density is the sum of a stellar disk, stellar bulge and gas component; while the first two are given in the SPARC catalogue, the third is not. We calculate it here from the SPARC $V_\text{gas}$ values using eq 13 of \citet{Toomre}. This requires an integral of $V_\text{gas}$ to $r = \infty$. We extrapolate it beyond the last measured point with a Keplerian decline, corresponding to the assumption of spherical symmetry and that all gas is enclosed within the observed region. To test the effect of this approximation, we consider an alternative (extreme) model in which $V_\text{gas}$ is constant beyond the last measured point, finding minimal difference in the results.

We use the MONDian $g$ calculated from Eq.~\ref{eq:bm} in place of Eq.~\ref{eq:g} in the likelihood term and repeat the inference for the $n$-family (which has simple analytic forms for both $\nu$ and $\mu$), without EFE. The results are shown in Table~\ref{tab:results_MG}. We see a decrease in the quality of the fit (shown by $\sig$ and $\Delta$BIC exceeding the corresponding rows of Table~\ref{tab:results_1} and~\ref{tab:results_2}) and an increased disagreement between the RAR, $Q_2$ and WBT constraints (shown by the reduced shape and increased $\sigma_{Q_2}$ and $\alpha_\text{grav}$). Our main results using the algebraic MOND model are therefore conservative. It is also interesting to note that the RAR seems to favour the straight algebraic MOND relation over this modified gravity correction, although not with high significance.

\subsection{Constraints from Cassini and comparison to the RAR and WBT}
\label{sec:results_q2}

\begin{figure}
  \centering
  \includegraphics[width=0.45\textwidth]{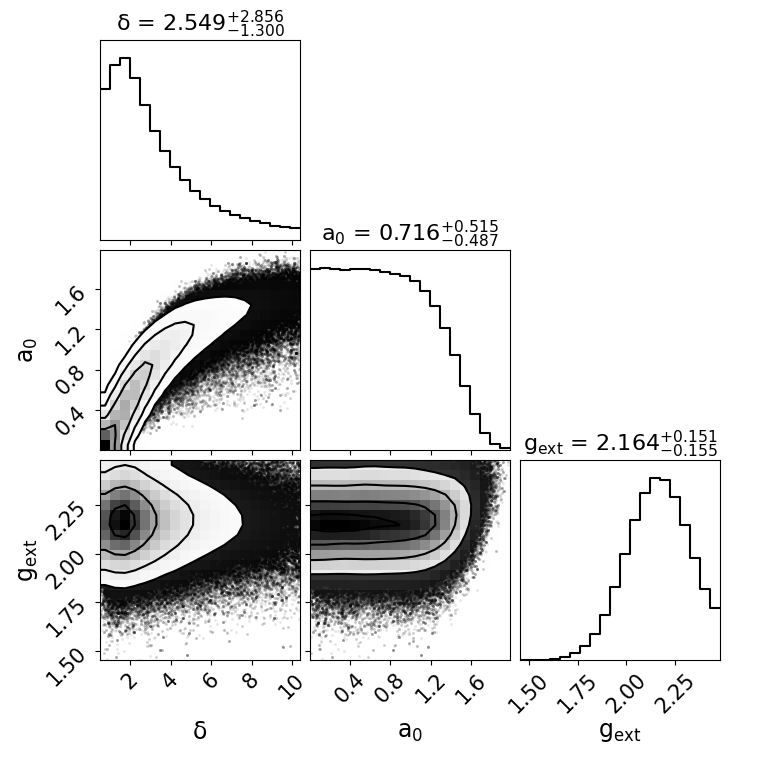}
  \caption{Constraints on $a_0$, $\delta$ and $g_{\rm ext}$ from the Cassini constraint $Q_2 = \left(3 \pm 3\right) \times 10^{-27}$ s$^{-2}$, using a flat prior on $Q_2$. $a_0$ and $g_\text{ext}$ are in units of $10^{-10}$ m s$^{-2}$. The $g_{\rm ext}$ posterior is dominated by its truncated Gaussian prior.}
  \label{fig:q2_corner}
\end{figure}

\begin{figure*}
  \centering
  \includegraphics[width=0.32\textwidth,height=0.33\textwidth]{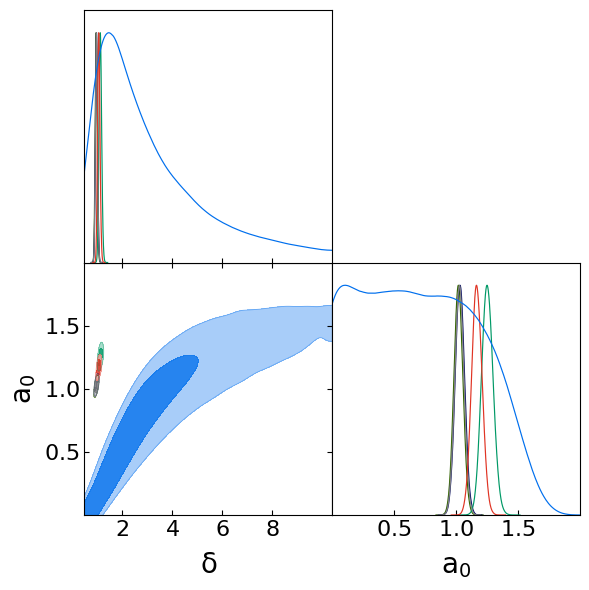}
  \includegraphics[width=0.32\textwidth,height=0.33\textwidth]{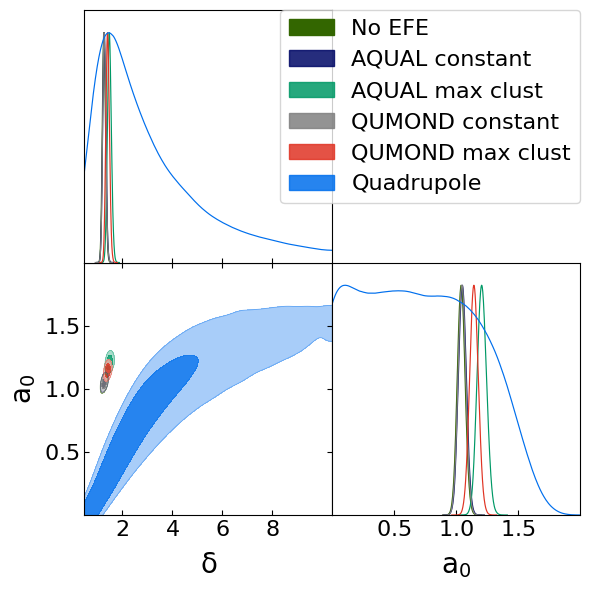}
  \includegraphics[width=0.32\textwidth,height=0.33\textwidth]{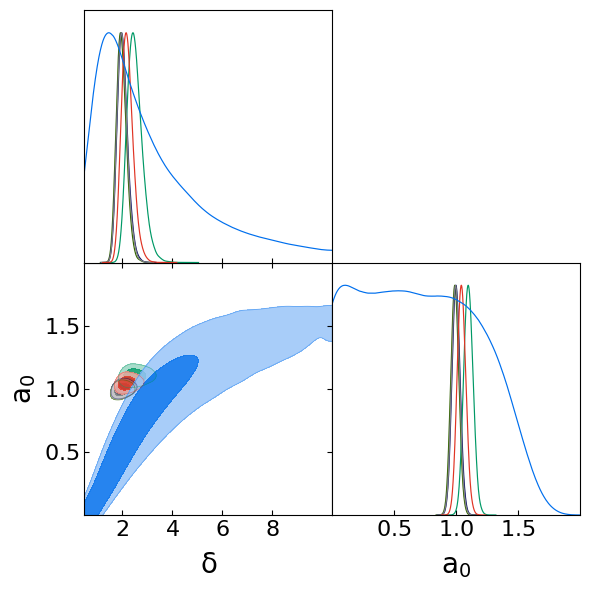}
  \caption{Posteriors on $a_0$ and $\delta$ from the RAR and SS quadrupole. \emph{Left:} Fiducial SPARC $M/L$ model; \emph{centre:} free Gaussian hyperpriors on $\Ud$ and $\Ub$; \emph{right:} as centre but excluding galaxies with bulges.}
  \label{fig:getdist}
\end{figure*}

To shed further light on the discrepancy at the heart of our analysis, we now infer
\{$a_0$, shape\} from the SS quadrupole. To set the stage, we show in Fig.~\ref{fig:q2_corner} the corner plot of the $Q_2$ inference for the $\delta$-family when $g_{\rm ext}$ is also inferred using a truncated Gaussian prior from \citet{klioner:2021} (see Sec.~\ref{sec:quadrupole_inference}). Higher shape and lower $a_0$ values are preferred by the likelihood because this pushes the SS further into the Newtonian regime where the prediction for $Q_2$ is 0. However, our flat prior on $Q_2$ creates a prior $\propto \text{d}(Q_2)/\text{d}(\text{shape})$ on shape, which goes to 0 at high shape as $Q_2$ levels out at 0. This truncates the posterior so that only a finite shape range is allowed. The shape and $a_0$ marginals only decline at small values (also in Fig.~\ref{fig:getdist}) due to our priors $\delta>0.5$ and $a_0>0.0025\times10^{-10}$ m s$^{-2}$, which do not impact the (in)consistency with the RAR.
There is little constraining power on $g_{\rm ext}$ from the $Q_2$ measurement, so its posterior is similar to its prior.

Next, in Fig.~\ref{fig:getdist} we compare the posteriors on $a_0$ and shape from the RAR and $Q_2$ inferences. In each panel we use the $\delta$-family and show the results for each of the EFE models; the three panels show the same $M/L$ models as in Figs.~\ref{fig:q2_ppd} and~\ref{fig:agrav_rar}.
We see again the clear tension between the RAR and $Q_2$ measurement using either the fiducial or Gaussian hyperprior $M/L$ model: the inferences only become consistent when the galaxies with bulges are removed.
This conclusion is not affected by the choice of IF family or EFE model.
If instead of sampling $g_{\rm ext}$ we choose the value in the range $2-2.48\times10^{-10}$m s$^{-2}$ that minimises $Q_2$ at given $a_0$ and shape (corresponding to the best possible $g_{\rm ext}$ for reducing the tension), the posterior is only slightly shifted to lower shape.

It is also instructive to consider the RAR fits obtained when the $Q_2$ constraint is enforced at the 1$\sigma$ level. To achieve this we fix $\delta=5$ and $a_0=1.45\times10^{-10}$ m s$^{-2}$---which yields $Q_2=6\times10^{-27}$ s$^{-2}$---and repeat the RAR inference using Gaussian hyperpriors on $\Ud$ and $\Ub$. For the no-EFE model, requiring this IF lowers $\ln(\hat{\mathcal{L}})$ by 195, indicating a hugely worse fit. The ln-distance prior is also worsened by 16.4 and the ln-inclination prior by 38.8, showing that these galaxy parameters are being forced to take values far from their priors.
The constraints on the $M/L$ hyperpriors are $\mu_\text{d}=0.92\pm0.03$ and $\mu_\text{b} = 0.67\pm0.04$, indicating an even greater deviation from the fiducial $M/L$ model than the regular Gaussian hyperprior case. $\sig$ is only 0.04 dex, however, and the plot of the transformed points with $\delta=5, a_0=1.45\times10^{-10}$m s$^{-2}$ model overlaid (not shown) looks similar to Fig.~\ref{fig:big_shape}, just with a somewhat sharper transition. Analogous results hold for the other EFE models.

\begin{figure}
  \centering
  \includegraphics[width=0.45\textwidth]{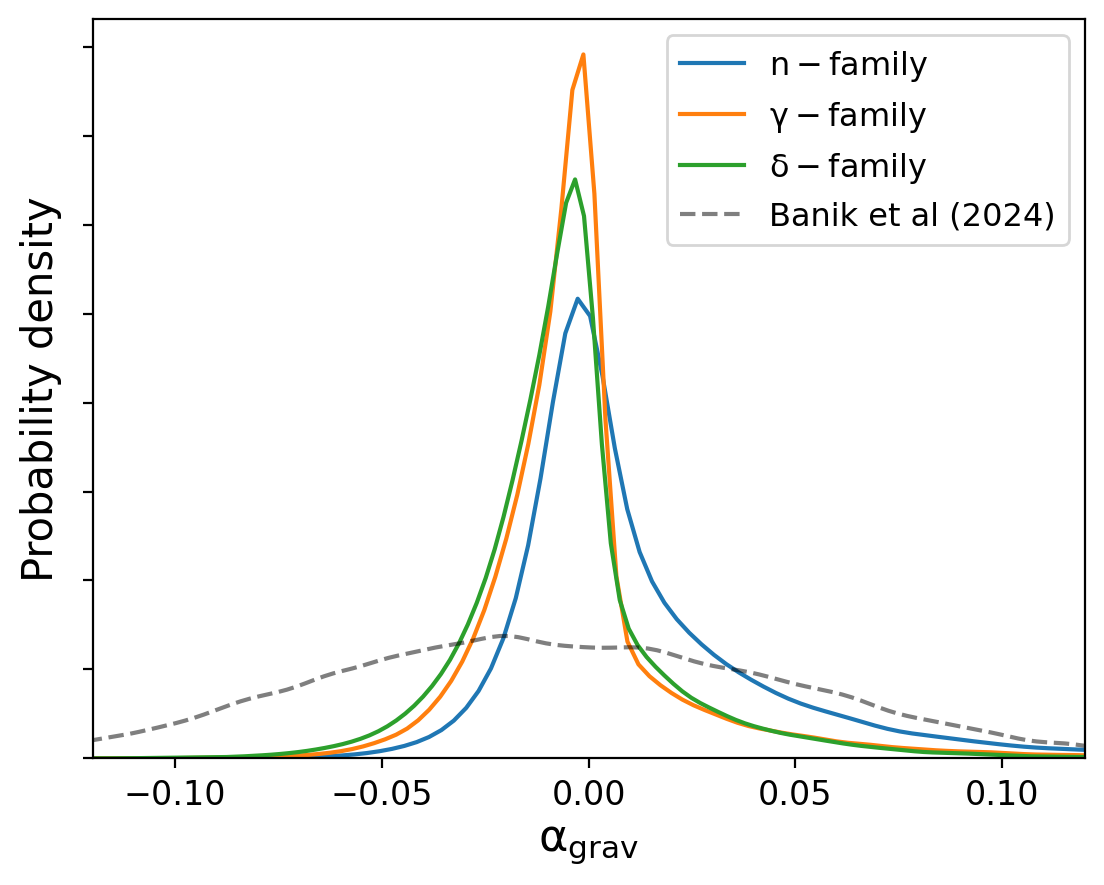}
  \caption{PPD of $\ag$ for the three IF families from the $Q_2$ constraint. This constraint foretells $\ag=0$ to higher precision than the WBT of~\citet{Banik_WBT}.}
  \label{fig:agrav_q2}
\end{figure}

Finally, we show in Fig.~\ref{fig:agrav_q2} the PPD of $\ag$ from the $Q_2$ chain.
Here we see that the Cassini measurement already imposes a stronger constraint on $\ag$ than does the WBT result of~\citet{Banik_WBT}: Eq.~\ref{eq:Q2_measurement} allows one to predict, from the SS alone, that the WBT will be null to high precision regardless of IF family.
This result may appear surprising given the fact that~\citeauthor{Banik_WBT} rule out the RAR IF with $a_0=1.2\times10^{-10}$ m s$^{-2}$ at 16$\sigma$, while we rule out the RAR IF using the Cassini measurement at only 8.7$\sigma$ (top row of Table~\ref{tab:results_1}).
This is primarily because $\ag$ is more sensitive to shape than is $Q_2$, so when allowing shape to increase from unity one achieves $\ag\approx0$ sooner than $Q_2\approx0$.
We therefore conclude that as a general constraint on modified gravity MOND, the SS quadrupole is stronger than the WBT that is currently possible. The PPD of $\ag$ from $Q_2$ agrees with the inference of~\citeauthor{Banik_WBT} but would not with those of~\citet{Hernandez, Hernandez_2, Hernandez_3, Hernandez_4, chae_wb_2, chae_wb}, who find approximate agreement with the Simple or RAR IF, $\ag\approx1$. Our results therefore argue in favour of~\citeauthor{Banik_WBT}.

\subsection{Can a fine-tuned IF reconcile the Cassini and RAR constraints?}
\label{sec:new_if}

One might be tempted to design a new IF with a sharp transition between the MONDian and Newtonian regimes in an attempt to reconcile the Cassini and RAR constraints, whilst still producing a deviation from Newtonian gravity in local wide binaries.
We have seen that removing bulges and adopting free hyperpriors for the stellar $M/L$ ratios of SPARC disks can reconcile the RAR and the Cassini constraints to better than 2$\sigma$. However, this produces no non-Newtonian behaviour in the WBT. Moreover, using fiducial SPARC $M/L$ values or including bulgey galaxies in the RAR fit returns us to a gradual transition close to the Simple or RAR IF.

In order to explore this question more systematically, let us now consider a class of IFs built from $\nu_{\rm RAR}$ by introducing a new acceleration scale $a_\mathrm{N,trans}$ where a sharp sigmoid transition is applied between $\nu_{\rm RAR}$ and the Newtonian regime ($\nu=1$). This means that for $y<a_\mathrm{N,trans}/a_0$ the IF behaves as $\nu_{\rm RAR}$, but it becomes fully Newtonian as soon as $y>a_\mathrm{N,trans}/a_0$. The case $a_\mathrm{N,trans}=a_0$ would correspond to a sharp transition from Newton to deep-MOND without introducing a new acceleration scale. However, to keep a positive WBT, one would need $a_\mathrm{N,trans}>a_0$ in order to produce non-Newtonian behaviour at the Sun's position in the Milky Way.

\begin{figure}
  \centering
  \includegraphics[width=0.5\textwidth]{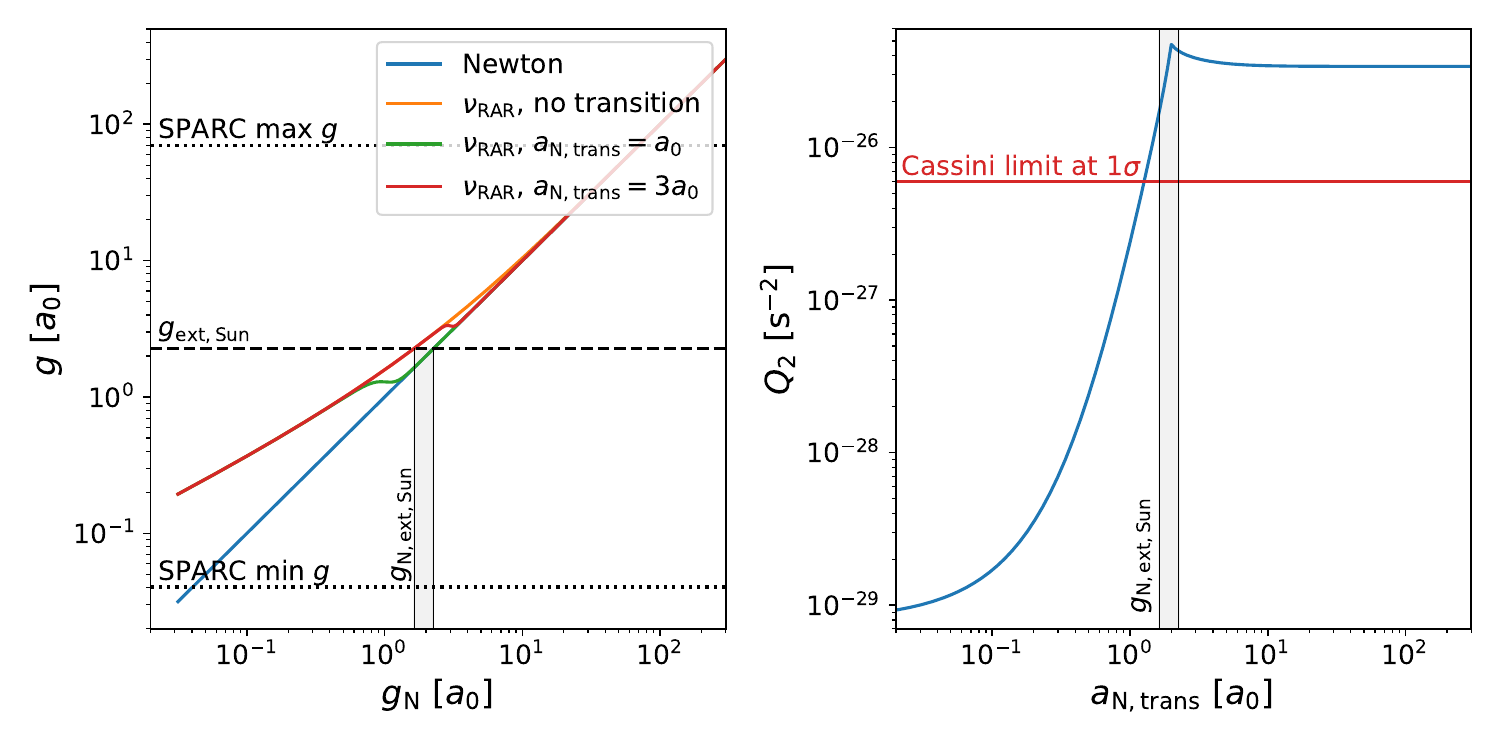}
  \caption{Exploration of a class of IFs constructed by introducing a sharp sigmoid transition between $\nu_{\rm RAR}$ and $\nu=1$ (fully Newtonian behaviour) at a new transition acceleration scale $a_\mathrm{N,trans}$. \emph{Left:} the RAR of this IF class for different values of $a_\mathrm{N,trans}$. The two horizontal dotted lines bound the gravitational fields probed by the SPARC RAR while the dashed horizontal line shows the external field acting on the SS, $g_\mathrm{ext,Sun}$. The vertical shaded area shows the range of Newtonian external fields corresponding to $g_\mathrm{ext,Sun}$ for different IFs. \emph{Right:} the variation of $Q_2$ produced by this class of IFs as a function of $a_\mathrm{N,trans}$. This transition must occur at an acceleration {\it below} the external field $g_\mathrm{N,ext,Sun}$ in order to satisfy the Cassini constraint at 1$\sigma$, which would both strongly impact the SPARC RAR and produce no non-Newtonian behaviour in the WBT (since local wide binaries are embedded in the same external field as the SS, where the IF would have to be fully Newtonian).}
  \label{fig:nu_transition}
\end{figure}

The left panel of Fig.~\ref{fig:nu_transition} shows the RAR obtained with this class of fine-tuned IFs for different values of $a_\mathrm{N,trans}$. We also display the minimum and maximum observed accelerations probed by the SPARC RCs (using the cuts defined in Sec.~\ref{sec:data_sparc}), and the external field of the Milky Way at the position of the sun (assuming $a_0=1.03\times 10^{-10}$ m s$^{-2}$; see Table~\ref{tab:results_1}). The right panel then shows the value of $Q_2$ in the SS as a function of $a_\mathrm{N,trans}$. It is clear that, in order to satisfy the Cassini constraint at 1$\sigma$, the new transition acceleration $a_\mathrm{N,trans}$ must be {\it below} the Newtonian external field at the Sun. This will both strongly affect the RAR fits, especially in bulgey galaxies that do not allow such a sharp transition as we have seen in Sec.~\ref{sec:ML_ext}, and necessarily yield no deviation from Newton in local wide binaries. It appears therefore to be highly unlikely that one could cook up a fine-tuned universal IF that would simultaneously explain the RAR and pass the Cassini constraint, especially if one also wants a non-Newtonian WBT.

\section{Discussion and Conclusion}
\label{sec:conc}

The simplicity and regularity of galaxy kinematics---epitomised for late-type galaxies' radial dynamics by the RAR---appears to provide strong support for MOND relative to the galaxy formation scenario of $\Lambda$CDM. We find excellent fits to the RAR of the SPARC galaxy sample using generalised IFs of the $n$-, $\delta$- or $\gamma$-families, with the very small intrinsic scatter ($\sim$8 per cent) approximately expected from remaining systematics or departures from spherical symmetry in modified gravity formulations of MOND. Including all galaxies, we find the best-fit shape of the RAR to be very close to the Simple IF or the one proposed in \citet{MLS}. The impact of the external field effect, while not clearly required in our analysis, is consistent with the expectation from independent calculations of the galaxies' large-scale environments.

It is interesting to compare our best-fit $a_0$ values with those in the literature. We recover $a_0 \approx 1\times10^{-10}\:\mathrm{m}\,\mathrm{s}^{-2}$ when not including the EFE or when the EFE is fitted as a global parameter, and $a_0 \approx 1.2 \times 10^{-10}\:\mathrm{m}\,\mathrm{s}^{-2}$ when the EFE is allowed to vary galaxy-by-galaxy with the maximum clustering prior. Historically, the first estimates of $a_0$---fitted with the Standard IF until circa 2005---gave values around $a_0=1.2 \times 10^{-10}\:\mathrm{m}\,\mathrm{s}^{-2}$ \citep[e.g.][]{Kent_a0,Milgrom_a0,Begeman}, typically using fewer than 10 RCs. Complementing the data of \citet{Begeman} with data from \citet{Gentile_2004} and using the Simple IF, \citet{Famaey_2007b} found $a_0=1.35 \times 10^{-10}\:\mathrm{m}\,\mathrm{s}^{-2}$ with 14 RCs, whilst a later analysis of 12 higher resolution RCs from the THINGS survey \citep{Gentile} gave $a_0=1.22 \times 10^{-10}\:\mathrm{m}\,\mathrm{s}^{-2}$ with the Simple IF, but with an intriguing tension for the archetypal RC of NGC~3198, indicating $a_0=0.9 \times 10^{-10}\:\mathrm{m}\,\mathrm{s}^{-2}$ in line with the more recent EFE-free estimates. Including the local EFE gives back the ``old'' value of $a_0 \approx 1.2 \times 10^{-10}\:\mathrm{m}\,\mathrm{s}^{-2}$.

Our main result is that the location and sharpness of the transition between the deep-MOND and Newtonian regimes inferred from the RAR is grossly inconsistent with that inferred from the Cassini measurement of the SS quadrupole in classical modified gravity formulations of MOND. Including full marginalisation over all relevant galaxy nuisance parameters we calculate this tension to be $8.7\sigma$ for the $\delta$-family of IFs and fiducial $M/L$ model of SPARC. We use the algebraic MOND relation between $\go$ and $\gb$ fiducially, but show explicitly that switching to an AQUAL fitting formula does not qualitatively affect the results. It is also interesting to note that the RAR seems to slightly favour the straight algebraic MOND relation over this modified gravity correction, although not with high significance.

We find that the tension can be slightly ameliorated by allowing the mass-to-light ratios $\Upsilon$ of the disks and bulges of the SPARC galaxies to float, which is strongly favoured by the RAR data (given an IF-family fit). This significantly increases the sharpness of the MOND transition---making the Solar neighbourhood more Newtonian---but prefers $\Ud>\Ub$, flying in the face of stellar population modelling. Using Gaussian hyperpriors with free means for $\Ud$ and $\Ub$ reduces the $Q_2$ tension to 7.2$\sigma$. A much greater gain is to be had by discarding the 31 galaxies with bulges, in which case the RAR with free $\Ud$ is in only 1.9$\sigma$ tension with the quadrupole constraint using the AQUAL galaxy-by-galaxy EFE model.
The resulting model cannot fit galaxies with bulges, however.
We stress that these modifications to the quadrupole prediction arise entirely internally to the RAR modelling and without any reference to the Cassini measurement: it is not a case of ameliorating tension in a joint inference by enlarging the parameter space, but rather a preference within the SPARC data.
We also show that, for the wide range of parametric RAR shapes considered here, the Cassini measurement implies (in the modified gravity context) no detectable deviation from Newtonian gravity in wide binaries in the Solar neighbourhood to greater precision than Gaia measurements of the wide binaries currently afford. This is consistent with
(and renders unsurprising) the results of~\citet{Banik_WBT}, but disagrees with \citet{Hernandez, Hernandez_2, Hernandez_3, Hernandez_4, chae_wb_2, chae_wb, Hernandez_Chae}.

Thus, while the RAR \emph{does} appear to be tantamount to a natural law in galaxies~\citep{OneLaw, uRAR, Stiskalek}, extending it into the Solar neighbourhood
does not work unless the SPARC $M/L$ model and/or treatment of bulges is greatly in error. How should we interpret this perplexing situation? Aside from the conclusion that the simplicity of galaxy dynamics is a red herring---a to-be-understood emergent behaviour of dark matter---and MOND on a hiding to nothing, there appear to be three possibilities:
\begin{enumerate}
\item Our different results from considering separately the bulgey and bulge-free galaxies---and preference for $\Upsilon$s far removed from the fiducial SPARC values---may be indicating that there are systematics in the RAR data which could permit a MOND-to-Newtonian transition as steep as is required by the SS quadrupole. Removing galaxies with bulges and allowing $\Ud$ to float freely is the most conservative model if one has reason to doubt the bulge modelling, and it may not be a coincidence that it is consistent with the Cassini measurement. In this case, the orbits of WBs in the Solar neighbourhood are also expected to be fully Newtonian.
This may however cause problems with observables not studied here, such as the RC of the Milky Way.

\hspace{6mm} There is no {\it a priori} reason to remove galaxies with bulges, but {\it a posteriori} the tension between galaxies with and without bulges provides strong motivation to examine this issue further (within the MOND paradigm).
The fact that the RAR \{$a_0$, shape\} constraints between bulgey and bulge-free galaxies and for various $M/L$ choices are clearly discrepant with one another (Fig.~\ref{fig:getdist_bulcomp}) is concerning in itself for the universality of the RAR. One may wonder whether the RAR parameters may correlate with other galaxy properties; this appears not to be the case from~\citet{Stiskalek}. Note that the bulgey versus bulge-free dichotomy is different to that of~\citet{Chae_distinguishing}, where it is instead between the RARs followed by inner and outer points of the RCs regardless of bulge fraction. This is unlikely to remain in the underlying RARs presented here due to their extremely small intrinsic scatter.
\item Remaining within the modified gravity paradigm and taking the inconsistency of the RAR and $Q_2$ results at face value necessitates a modification to the AQUAL or QUMOND weak-field limits of any such theory. One way to achieve this would be to screen the modification to GR on small scales, a method already common for preventing scalar--tensor theories of gravity from manifesting unobserved deviations from the inverse square law (``fifth forces'') in the SS (see~\citealt{Jain_Khoury,NPP} for reviews). This may be achieved in the case of MOND by, for instance, including a covariant Galileon-type term (involving a length-scale) for the scalar field in the MOND action \citep{Babichev}: this could completely suppress all MOND effects below a Vainshtein radius depending on the source mass, $a_0$ and the new length-scale. Interestingly, with the length-scale proposed by \citet{Babichev}, MOND effects would be killed in the SS and up to the typical (mass and length) scale of WBs. Such a Galileon term could therefore be added to frameworks such as those of \citet{skordis}, and would not affect the predicted RAR on larger scales where interesting deviations from the algebraic MOND relation are already to be found. This is studied for  high-mass (galaxy cluster-like) systems in \citet{Durakovic}; see also~\citet{Banik_WBT}. The idea of adding a new scale to MOND is somewhat akin to the approach of ``Extended MOND'' where a second gravitational variable (in this case potential) modulates the MOND boost~\citep{emond}.

\hspace{6mm} It is important to bear in mind that our conclusions refer solely to specific modified gravity instantiations of MOND in the classical regime,
namely the AQUAL and QUMOND models where the IF is necessarily universal. Generalisations such as TRIMOND \citep{TRIMOND} and GQUMOND \citep{GQUMOND} have more complex behaviour, and may evade our constraints by having different IFs in galaxies, wide binaries and the SS. In GQUMOND a new scale naturally arises, allowing MOND effects to be screened below some length-scale (as in the Galileon screening case) or dynamical time. TRIMOND manifests three gravitational degrees of freedom---the MOND potential and two auxiliary potentials, one of which is the Newtonian potential---and contains AQUAL and QUMOND as special cases \citep{TRIMOND}. There thus remains an extensive theory space of MOND-like theories, which
could be constrained through the apparent incompatibility revealed here. However, one may consider that we are reaching the point where the complexity of modified gravity formulations of MOND is no longer warranted by its empirical successes relative to $\Lambda$CDM.
\item Finally, one may consider MOND as implying a modification to inertia. Modified inertia formulations are far more difficult to construct and study because the dynamics of an object may depend on its entire past history (e.g.~\citealt{Milgrom_2011,inertia_1}). It is however modified inertia that produces in its pristine form the strongest piece of MOND phenomenology, viz. the RAR~\citep{MI_1994, inertia_1}. In this context, the mismatch between bulgey and bulge-free galaxies could be a manifestation of the subtle dependence of the MOND force law on the underlying orbital dynamics and history. Our results may therefore be indicating that modified inertia is the correct interpretation of MOND, spurring the search for theories and models that allow it to make contact with data beyond disc galaxy dynamics.
Indeed, a Solar System quadrupole is not generically predicted by modified inertia~\citet{MI_2023}, potentially allowing the Cassini constraint to be completely circumvented.
\end{enumerate}

\section*{Acknowledgements}

We thank Indranil Banik, Michal Bilek, Amel Durakovic, Federico Lelli, Stacy McGaugh, Mordehai Milgrom, Constantinos Skordis, Richard Stiskalek and Tariq Yasin for useful discussions.

HD is supported by a Royal Society University Research Fellowship (grant no. 211046).
BF acknowledges funding from the European Research Council (ERC) under the European Union's Horizon 2020 research and innovation program (grant agreement No. 834148) and from the Agence Nationale de la Recherche (ANR projects ANR-18-CE31-0006 and ANR-19-CE31-0017).

This project has received funding from the European Research Council (ERC) under the European Union's Horizon 2020 research and innovation programme (grant agreement No. 693024).

For the purpose of open access, the authors have applied a Creative Commons Attribution (CC BY) licence to any Author Accepted Manuscript version arising.

\section*{Data availability}

All data underlying this article may be shared on request to the authors.

\bibliographystyle{mnras}
\bibliography{sparc}

\begin{thebibliography}{}
\makeatletter
\relax
\def\mn@urlcharsother{\let\do\@makeother \do\$\do\&\do\#\do\^\do\_\do\%\do\~}
\def\mn@doi{\begingroup\mn@urlcharsother \@ifnextchar [ {\mn@doi@}
  {\mn@doi@[]}}
\def\mn@doi@[#1]#2{\def\@tempa{#1}\ifx\@tempa\@empty \href
  {http://dx.doi.org/#2} {doi:#2}\else \href {http://dx.doi.org/#2} {#1}\fi
  \endgroup}
\def\mn@eprint#1#2{\mn@eprint@#1:#2::\@nil}
\def\mn@eprint@arXiv#1{\href {http://arxiv.org/abs/#1} {{\tt arXiv:#1}}}
\def\mn@eprint@dblp#1{\href {http://dblp.uni-trier.de/rec/bibtex/#1.xml}
  {dblp:#1}}
\def\mn@eprint@#1:#2:#3:#4\@nil{\def\@tempa {#1}\def\@tempb {#2}\def\@tempc
  {#3}\ifx \@tempc \@empty \let \@tempc \@tempb \let \@tempb \@tempa \fi \ifx
  \@tempb \@empty \def\@tempb {arXiv}\fi \@ifundefined
  {mn@eprint@\@tempb}{\@tempb:\@tempc}{\expandafter \expandafter \csname
  mn@eprint@\@tempb\endcsname \expandafter{\@tempc}}}

\bibitem[\protect\citeauthoryear{{Adelberger}, {Heckel}  \&
  {Nelson}}{{Adelberger} et~al.}{2003}]{Adelberger}
{Adelberger} E.~G.,  {Heckel} B.~R.,   {Nelson} A.~E.,  2003, \mn@doi [Annual
  Review of Nuclear and Particle Science]
  {10.1146/annurev.nucl.53.041002.110503}, \href
  {https://ui.adsabs.harvard.edu/abs/2003ARNPS..53...77A} {53, 77}

\bibitem[\protect\citeauthoryear{{Angus}, {Famaey}  \& {Buote}}{{Angus}
  et~al.}{2008}]{clusters_4}
{Angus} G.~W.,  {Famaey} B.,   {Buote} D.~A.,  2008, \mn@doi [\mnras]
  {10.1111/j.1365-2966.2008.13353.x}, \href
  {https://ui.adsabs.harvard.edu/abs/2008MNRAS.387.1470A} {387, 1470}

\bibitem[\protect\citeauthoryear{{Antreasian} et~al.,}{{Antreasian}
  et~al.}{2005}]{antreasian:2005kx}
{Antreasian} P.~G.,  et~al., 2005, in AAS/AIAA Astrodynamics Conference.
  Pasadena, CA : Jet Propulsion Laboratory, National Aeronautics and Space
  Administration, 2005., pp AAS 05--312

\bibitem[\protect\citeauthoryear{{Asencio}, {Banik}, {Mieske}, {Venhola},
  {Kroupa}  \& {Zhao}}{{Asencio} et~al.}{2022}]{Asencio}
{Asencio} E.,  {Banik} I.,  {Mieske} S.,  {Venhola} A.,  {Kroupa} P.,   {Zhao}
  H.,  2022, \mn@doi [\mnras] {10.1093/mnras/stac1765}, \href
  {https://ui.adsabs.harvard.edu/abs/2022MNRAS.515.2981A} {515, 2981}

\bibitem[\protect\citeauthoryear{{Babichev}, {Deffayet}  \&
  {Esposito-Far{\`e}se}}{{Babichev} et~al.}{2011}]{Babichev}
{Babichev} E.,  {Deffayet} C.,   {Esposito-Far{\`e}se} G.,  2011, \mn@doi
  [\prd] {10.1103/PhysRevD.84.061502}, \href
  {https://ui.adsabs.harvard.edu/abs/2011PhRvD..84f1502B} {84, 061502}

\bibitem[\protect\citeauthoryear{Baker et~al.}{Baker et~al.}{2021}]{NPP}
Baker T.,  et~al., 2021, \mn@doi [Rev. Mod. Phys.]
  {10.1103/RevModPhys.93.015003}, 93, 015003

\bibitem[\protect\citeauthoryear{{Banik} \& {Zhao}}{{Banik} \&
  {Zhao}}{2022}]{Banik}
{Banik} I.,  {Zhao} H.,  2022, \mn@doi [Symmetry] {10.3390/sym14071331}, \href
  {https://ui.adsabs.harvard.edu/abs/2022Symm...14.1331B} {14, 1331}

\bibitem[\protect\citeauthoryear{{Banik}, {Thies}, {Famaey}, {Candlish},
  {Kroupa}  \& {Ibata}}{{Banik} et~al.}{2020}]{Banik_M33}
{Banik} I.,  {Thies} I.,  {Famaey} B.,  {Candlish} G.,  {Kroupa} P.,   {Ibata}
  R.,  2020, \mn@doi [\apj] {10.3847/1538-4357/abc623}, \href
  {https://ui.adsabs.harvard.edu/abs/2020ApJ...905..135B} {905, 135}

\bibitem[\protect\citeauthoryear{{Banik}, {Pittordis}, {Sutherland}, {Famaey},
  {Ibata}, {Mieske}  \& {Zhao}}{{Banik} et~al.}{2024}]{Banik_WBT}
{Banik} I.,  {Pittordis} C.,  {Sutherland} W.,  {Famaey} B.,  {Ibata} R.,
  {Mieske} S.,   {Zhao} H.,  2024, \mn@doi [\mnras] {10.1093/mnras/stad3393},
  \href {https://ui.adsabs.harvard.edu/abs/2024MNRAS.527.4573B} {527, 4573}

\bibitem[\protect\citeauthoryear{{Begeman}, {Broeils}  \& {Sanders}}{{Begeman}
  et~al.}{1991}]{Begeman}
{Begeman} K.~G.,  {Broeils} A.~H.,   {Sanders} R.~H.,  1991, \mn@doi [\mnras]
  {10.1093/mnras/249.3.523}, \href
  {https://ui.adsabs.harvard.edu/abs/1991MNRAS.249..523B} {249, 523}

\bibitem[\protect\citeauthoryear{{Bekenstein} \& {Milgrom}}{{Bekenstein} \&
  {Milgrom}}{1984}]{aqual}
{Bekenstein} J.,  {Milgrom} M.,  1984, \mn@doi [\apj] {10.1086/162570}, \href
  {https://ui.adsabs.harvard.edu/abs/1984ApJ...286....7B} {286, 7}

\bibitem[\protect\citeauthoryear{Bingham et~al.,}{Bingham
  et~al.}{2019}]{bingham2019pyro}
Bingham E.,  et~al., 2019, J. Mach. Learn. Res., 20, 28:1

\bibitem[\protect\citeauthoryear{{Blanchet} \& {Novak}}{{Blanchet} \&
  {Novak}}{2011}]{blanchet:2011ys}
{Blanchet} L.,  {Novak} J.,  2011, \mn@doi [\mnras]
  {10.1111/j.1365-2966.2010.18076.x}, 412, 2530

\bibitem[\protect\citeauthoryear{{Bovy}, {Bird}, {Garc{\'\i}a P{\'e}rez},
  {Majewski}, {Nidever}  \& {Zasowski}}{{Bovy} et~al.}{2015}]{Bovy}
{Bovy} J.,  {Bird} J.~C.,  {Garc{\'\i}a P{\'e}rez} A.~E.,  {Majewski} S.~R.,
  {Nidever} D.~L.,   {Zasowski} G.,  2015, \mn@doi [\apj]
  {10.1088/0004-637X/800/2/83}, \href
  {https://ui.adsabs.harvard.edu/abs/2015ApJ...800...83B} {800, 83}

\bibitem[\protect\citeauthoryear{Brada \& Milgrom}{Brada \&
  Milgrom}{1995}]{Brada_Milgrom}
Brada R.,  Milgrom M.,  1995, \mn@doi [Mon. Not. Roy. Astron. Soc.]
  {10.1093/mnras/276.2.453}, 276, 453

\bibitem[\protect\citeauthoryear{{Brown} \& {Mathur}}{{Brown} \&
  {Mathur}}{2023}]{Brown}
{Brown} K.,  {Mathur} H.,  2023, \mn@doi [\aj] {10.3847/1538-3881/acef1e},
  \href {https://ui.adsabs.harvard.edu/abs/2023AJ....166..168B} {166, 168}

\bibitem[\protect\citeauthoryear{Brown, Abraham, Kell  \& Mathur}{Brown
  et~al.}{2018}]{Jones-Smith}
Brown K.,  Abraham R.,  Kell L.,   Mathur H.,  2018, \mn@doi [New Journal of
  Physics] {10.1088/1367-2630/aaca23}, 20, 063042

\bibitem[\protect\citeauthoryear{{Chae}}{{Chae}}{2022}]{Chae_distinguishing}
{Chae} K.-H.,  2022, \mn@doi [\apj] {10.3847/1538-4357/ac93fc}, \href
  {https://ui.adsabs.harvard.edu/abs/2022ApJ...941...55C} {941, 55}

\bibitem[\protect\citeauthoryear{{Chae}}{{Chae}}{2023}]{chae_wb}
{Chae} K.-H.,  2023, \mn@doi [\apj] {10.3847/1538-4357/ace101}, \href
  {https://ui.adsabs.harvard.edu/abs/2023ApJ...952..128C} {952, 128}

\bibitem[\protect\citeauthoryear{{Chae}}{{Chae}}{2024}]{chae_wb_2}
{Chae} K.-H.,  2024, \mn@doi [\apj] {10.3847/1538-4357/ad0ed5}, \href
  {https://ui.adsabs.harvard.edu/abs/2024ApJ...960..114C} {960, 114}

\bibitem[\protect\citeauthoryear{{Chae} \& {Milgrom}}{{Chae} \&
  {Milgrom}}{2022}]{chae:2022tt}
{Chae} K.-H.,  {Milgrom} M.,  2022, \mn@doi [\apj] {10.3847/1538-4357/ac5405},
  \href {https://ui.adsabs.harvard.edu/abs/2022ApJ...928...24C} {928, 24}

\bibitem[\protect\citeauthoryear{{Chae}, {Lelli}, {Desmond}, {McGaugh}, {Li}
  \& {Schombert}}{{Chae} et~al.}{2020a}]{Paper_I}
{Chae} K.-H.,  {Lelli} F.,  {Desmond} H.,  {McGaugh} S.~S.,  {Li} P.,
  {Schombert} J.~M.,  2020a, \mn@doi [\apj] {10.3847/1538-4357/abbb96}, \href
  {https://ui.adsabs.harvard.edu/abs/2020ApJ...904...51C} {904, 51}

\bibitem[\protect\citeauthoryear{{Chae}, {Lelli}, {Desmond}, {McGaugh}, {Li}
  \& {Schombert}}{{Chae} et~al.}{2020b}]{Chae_1}
{Chae} K.-H.,  {Lelli} F.,  {Desmond} H.,  {McGaugh} S.~S.,  {Li} P.,
  {Schombert} J.~M.,  2020b, \mn@doi [\apj] {10.3847/1538-4357/abbb96}, \href
  {https://ui.adsabs.harvard.edu/abs/2020ApJ...904...51C} {904, 51}

\bibitem[\protect\citeauthoryear{{Chae}, {Desmond}, {Lelli}, {McGaugh}  \&
  {Schombert}}{{Chae} et~al.}{2021}]{Chae_2}
{Chae} K.-H.,  {Desmond} H.,  {Lelli} F.,  {McGaugh} S.~S.,   {Schombert}
  J.~M.,  2021, \mn@doi [\apj] {10.3847/1538-4357/ac1bba}, \href
  {https://ui.adsabs.harvard.edu/abs/2021ApJ...921..104C} {921, 104}

\bibitem[\protect\citeauthoryear{{Chae}, {Lelli}, {Desmond}, {McGaugh}  \&
  {Schombert}}{{Chae} et~al.}{2022}]{Chae_3}
{Chae} K.-H.,  {Lelli} F.,  {Desmond} H.,  {McGaugh} S.~S.,   {Schombert}
  J.~M.,  2022, \mn@doi [\prd] {10.1103/PhysRevD.106.103025}, \href
  {https://ui.adsabs.harvard.edu/abs/2022PhRvD.106j3025C} {106, 103025}

\bibitem[\protect\citeauthoryear{{Desmond}}{{Desmond}}{2017a}]{Desmond_MDAR}
{Desmond} H.,  2017a, \mn@doi [\mnras] {10.1093/mnras/stw2571}, \href
  {http://adsabs.harvard.edu/abs/2017MNRAS.464.4160D} {464, 4160}

\bibitem[\protect\citeauthoryear{{Desmond}}{{Desmond}}{2017b}]{Desmond_BTFR}
{Desmond} H.,  2017b, \mn@doi [\mnras] {10.1093/mnrasl/slx134}, \href
  {https://ui.adsabs.harvard.edu/abs/2017MNRAS.472L..35D} {472, L35}

\bibitem[\protect\citeauthoryear{{Desmond}}{{Desmond}}{2023}]{uRAR}
{Desmond} H.,  2023, \mn@doi [\mnras] {10.1093/mnras/stad2762}, \href
  {https://ui.adsabs.harvard.edu/abs/2023MNRAS.526.3342D} {526, 3342}

\bibitem[\protect\citeauthoryear{{Desmond} \& {Wechsler}}{{Desmond} \&
  {Wechsler}}{2015}]{Desmond_TFR}
{Desmond} H.,  {Wechsler} R.~H.,  2015, \mn@doi [\mnras]
  {10.1093/mnras/stv1978}, \href
  {https://ui.adsabs.harvard.edu/abs/2015MNRAS.454..322D} {454, 322}

\bibitem[\protect\citeauthoryear{{Desmond}, {Katz}, {Lelli}  \&
  {McGaugh}}{{Desmond} et~al.}{2019}]{Desmond_uncorrelated}
{Desmond} H.,  {Katz} H.,  {Lelli} F.,   {McGaugh} S.,  2019, \mn@doi [\mnras]
  {10.1093/mnras/stz016}, \href
  {https://ui.adsabs.harvard.edu/abs/2019MNRAS.484..239D} {484, 239}

\bibitem[\protect\citeauthoryear{{Desmond}, {Bartlett}  \&
  {Ferreira}}{{Desmond} et~al.}{2023}]{ESR-RAR}
{Desmond} H.,  {Bartlett} D.~J.,   {Ferreira} P.~G.,  2023, \mn@doi [\mnras]
  {10.1093/mnras/stad597}, \href
  {https://ui.adsabs.harvard.edu/abs/2023MNRAS.521.1817D} {521, 1817}

\bibitem[\protect\citeauthoryear{{Di Cintio} \& {Lelli}}{{Di Cintio} \&
  {Lelli}}{2016}]{DC_Lelli}
{Di Cintio} A.,  {Lelli} F.,  2016, \mn@doi [\mnras] {10.1093/mnrasl/slv185},
  \href {http://adsabs.harvard.edu/abs/2016MNRAS.456L.127D} {456, L127}

\bibitem[\protect\citeauthoryear{{Durakovic} \& {Skordis}}{{Durakovic} \&
  {Skordis}}{2023}]{Durakovic}
{Durakovic} A.,  {Skordis} C.,  2023, \mn@doi [arXiv e-prints]
  {10.48550/arXiv.2312.00889}, \href
  {https://ui.adsabs.harvard.edu/abs/2023arXiv231200889D} {p. arXiv:2312.00889}

\bibitem[\protect\citeauthoryear{{Famaey} \& {Binney}}{{Famaey} \&
  {Binney}}{2005}]{Simple_IF}
{Famaey} B.,  {Binney} J.,  2005, \mn@doi [\mnras]
  {10.1111/j.1365-2966.2005.09474.x}, \href
  {https://ui.adsabs.harvard.edu/abs/2005MNRAS.363..603F} {363, 603}

\bibitem[\protect\citeauthoryear{{Famaey} \& {McGaugh}}{{Famaey} \&
  {McGaugh}}{2012}]{FamaeyMcG}
{Famaey} B.,  {McGaugh} S.~S.,  2012, \mn@doi [Living Reviews in Relativity]
  {10.12942/lrr-2012-10}, \href
  {http://adsabs.harvard.edu/abs/2012LRR....15...10F} {15, 10}

\bibitem[\protect\citeauthoryear{{Famaey}, {Gentile}, {Bruneton}  \&
  {Zhao}}{{Famaey} et~al.}{2007a}]{Famaey_2007b}
{Famaey} B.,  {Gentile} G.,  {Bruneton} J.-P.,   {Zhao} H.,  2007a, \mn@doi
  [\prd] {10.1103/PhysRevD.75.063002}, \href
  {https://ui.adsabs.harvard.edu/abs/2007PhRvD..75f3002F} {75, 063002}

\bibitem[\protect\citeauthoryear{{Famaey}, {Bruneton}  \& {Zhao}}{{Famaey}
  et~al.}{2007b}]{Famaey_2007a}
{Famaey} B.,  {Bruneton} J.-P.,   {Zhao} H.,  2007b, \mn@doi [\mnras]
  {10.1111/j.1745-3933.2007.00308.x}, \href
  {https://ui.adsabs.harvard.edu/abs/2007MNRAS.377L..79F} {377, L79}

\bibitem[\protect\citeauthoryear{{Famaey}, {McGaugh}  \& {Milgrom}}{{Famaey}
  et~al.}{2018}]{Famaey_2018}
{Famaey} B.,  {McGaugh} S.,   {Milgrom} M.,  2018, \mn@doi [\mnras]
  {10.1093/mnras/sty1884}, \href
  {https://ui.adsabs.harvard.edu/abs/2018MNRAS.480..473F} {480, 473}

\bibitem[\protect\citeauthoryear{{Folkner}, {Williams}, {Boggs}, {Park}  \&
  {Kuchynka}}{{Folkner} et~al.}{2014}]{folkner:2014uq}
{Folkner} W.~M.,  {Williams} J.~G.,  {Boggs} D.~H.,  {Park} R.,   {Kuchynka}
  P.,  2014, IPN Progress Report, 42

\bibitem[\protect\citeauthoryear{{Freundlich}, {Famaey}, {Oria}, {B{\'\i}lek},
  {M{\"u}ller}  \& {Ibata}}{{Freundlich} et~al.}{2022}]{Freundlich}
{Freundlich} J.,  {Famaey} B.,  {Oria} P.-A.,  {B{\'\i}lek} M.,  {M{\"u}ller}
  O.,   {Ibata} R.,  2022, \mn@doi [\aap] {10.1051/0004-6361/202142060}, \href
  {https://ui.adsabs.harvard.edu/abs/2022A&A...658A..26F} {658, A26}

\bibitem[\protect\citeauthoryear{{Gaia Collaboration} et~al.,}{{Gaia
  Collaboration} et~al.}{2021}]{klioner:2021}
{Gaia Collaboration} et~al., 2021, \mn@doi [\aap]
  {10.1051/0004-6361/202039734}, \href
  {https://ui.adsabs.harvard.edu/abs/2021A&A...649A...9G} {649, A9}

\bibitem[\protect\citeauthoryear{{Gentile}, {Salucci}, {Klein}, {Vergani}  \&
  {Kalberla}}{{Gentile} et~al.}{2004}]{Gentile_2004}
{Gentile} G.,  {Salucci} P.,  {Klein} U.,  {Vergani} D.,   {Kalberla} P.,
  2004, \mn@doi [\mnras] {10.1111/j.1365-2966.2004.07836.x}, \href
  {https://ui.adsabs.harvard.edu/abs/2004MNRAS.351..903G} {351, 903}

\bibitem[\protect\citeauthoryear{{Gentile}, {Famaey}  \& {de Blok}}{{Gentile}
  et~al.}{2011}]{Gentile}
{Gentile} G.,  {Famaey} B.,   {de Blok} W.~J.~G.,  2011, \mn@doi [\aap]
  {10.1051/0004-6361/201015283}, \href
  {https://ui.adsabs.harvard.edu/abs/2011A&A...527A..76G} {527, A76}

\bibitem[\protect\citeauthoryear{{Ghari}, {Haghi}  \& {Zonoozi}}{{Ghari}
  et~al.}{2019a}]{gas_scaling}
{Ghari} A.,  {Haghi} H.,   {Zonoozi} A.~H.,  2019a, \mn@doi [\mnras]
  {10.1093/mnras/stz1272}, \href
  {https://ui.adsabs.harvard.edu/abs/2019MNRAS.487.2148G} {487, 2148}

\bibitem[\protect\citeauthoryear{{Ghari}, {Famaey}, {Laporte}  \&
  {Haghi}}{{Ghari} et~al.}{2019b}]{Ghari}
{Ghari} A.,  {Famaey} B.,  {Laporte} C.,   {Haghi} H.,  2019b, \mn@doi [\aap]
  {10.1051/0004-6361/201834661}, \href
  {https://ui.adsabs.harvard.edu/abs/2019A&A...623A.123G} {623, A123}

\bibitem[\protect\citeauthoryear{{Haghi}, {Baumgardt}, {Kroupa}, {Grebel},
  {Hilker}  \& {Jordi}}{{Haghi} et~al.}{2009}]{Haghi}
{Haghi} H.,  {Baumgardt} H.,  {Kroupa} P.,  {Grebel} E.~K.,  {Hilker} M.,
  {Jordi} K.,  2009, \mn@doi [\mnras] {10.1111/j.1365-2966.2009.14656.x}, \href
  {http://adsabs.harvard.edu/abs/2009MNRAS.395.1549H} {395, 1549}

\bibitem[\protect\citeauthoryear{{Haghi}, {Bazkiaei}, {Zonoozi}  \&
  {Kroupa}}{{Haghi} et~al.}{2016}]{Haghi_2016}
{Haghi} H.,  {Bazkiaei} A.~E.,  {Zonoozi} A.~H.,   {Kroupa} P.,  2016, \mn@doi
  [\mnras] {10.1093/mnras/stw573}, \href
  {https://ui.adsabs.harvard.edu/abs/2016MNRAS.458.4172H} {458, 4172}

\bibitem[\protect\citeauthoryear{{Haghi} et~al.,}{{Haghi}
  et~al.}{2019}]{Haghi_2019}
{Haghi} H.,  et~al., 2019, \mn@doi [\mnras] {10.1093/mnras/stz1465}, \href
  {https://ui.adsabs.harvard.edu/abs/2019MNRAS.487.2441H} {487, 2441}

\bibitem[\protect\citeauthoryear{{Hees} et~al.,}{{Hees}
  et~al.}{2012}]{hees:2012fk}
{Hees} A.,  et~al., 2012, \mn@doi [{Classical and Quantum Gravity}]
  {10.1088/0264-9381/29/23/235027}, 29, 235027

\bibitem[\protect\citeauthoryear{{Hees}, {Folkner}, {Jacobson}  \&
  {Park}}{{Hees} et~al.}{2014}]{hees:2014jk}
{Hees} A.,  {Folkner} W.~M.,  {Jacobson} R.~A.,   {Park} R.~S.,  2014, \mn@doi
  [\prd] {10.1103/PhysRevD.89.102002}, \href
  {http://adsabs.harvard.edu/abs/2014PhRvD..89j2002H} {89, 102002}

\bibitem[\protect\citeauthoryear{{Hees}, {Famaey}, {Angus}  \&
  {Gentile}}{{Hees} et~al.}{2016}]{Hees_main}
{Hees} A.,  {Famaey} B.,  {Angus} G.~W.,   {Gentile} G.,  2016, \mn@doi
  [\mnras] {10.1093/mnras/stv2330}, \href
  {https://ui.adsabs.harvard.edu/abs/2016MNRAS.455..449H} {455, 449}

\bibitem[\protect\citeauthoryear{{Hernandez}}{{Hernandez}}{2023}]{Hernandez_3}
{Hernandez} X.,  2023, \mn@doi [\mnras] {10.1093/mnras/stad2306}, \href
  {https://ui.adsabs.harvard.edu/abs/2023MNRAS.525.1401H} {525, 1401}

\bibitem[\protect\citeauthoryear{{Hernandez} \& {Chae}}{{Hernandez} \&
  {Chae}}{2023}]{Hernandez_Chae}
{Hernandez} X.,  {Chae} K.-H.,  2023, \mn@doi [arXiv e-prints]
  {10.48550/arXiv.2312.03162}, \href
  {https://ui.adsabs.harvard.edu/abs/2023arXiv231203162H} {p. arXiv:2312.03162}

\bibitem[\protect\citeauthoryear{{Hernandez}, {Cort{\'e}s}, {Allen}  \&
  {Scarpa}}{{Hernandez} et~al.}{2019}]{Hernandez}
{Hernandez} X.,  {Cort{\'e}s} R.~A.~M.,  {Allen} C.,   {Scarpa} R.,  2019,
  \mn@doi [International Journal of Modern Physics D]
  {10.1142/S0218271819501013}, \href
  {https://ui.adsabs.harvard.edu/abs/2019IJMPD..2850101H} {28, 1950101}

\bibitem[\protect\citeauthoryear{{Hernandez}, {Cookson}  \&
  {Cort{\'e}s}}{{Hernandez} et~al.}{2022}]{Hernandez_2}
{Hernandez} X.,  {Cookson} S.,   {Cort{\'e}s} R.~A.~M.,  2022, \mn@doi [\mnras]
  {10.1093/mnras/stab3038}, \href
  {https://ui.adsabs.harvard.edu/abs/2022MNRAS.509.2304H} {509, 2304}

\bibitem[\protect\citeauthoryear{Hernandez, Verteletskyi, Nasser  \&
  Aguayo-Ortiz}{Hernandez et~al.}{2023}]{Hernandez_4}
Hernandez X.,  Verteletskyi V.,  Nasser L.,   Aguayo-Ortiz A.,  2023, \mn@doi
  [\mnras] {10.1093/mnras/stad3446}, 528, 4720

\bibitem[\protect\citeauthoryear{{Iess}, {Rappaport}, {Jacobson}, {Racioppa},
  {Stevenson}, {Tortora}, {Armstrong}  \& {Asmar}}{{Iess}
  et~al.}{2010}]{iess:2010ys}
{Iess} L.,  {Rappaport} N.~J.,  {Jacobson} R.~A.,  {Racioppa} P.,  {Stevenson}
  D.~J.,  {Tortora} P.,  {Armstrong} J.~W.,   {Asmar} S.~W.,  2010, \mn@doi
  [Science] {10.1126/science.1182583}, \href
  {http://adsabs.harvard.edu/abs/2010Sci...327.1367I} {327, 1367}

\bibitem[\protect\citeauthoryear{{Iess} et~al.,}{{Iess}
  et~al.}{2012}]{iess:2012vn}
{Iess} L.,  et~al., 2012, \mn@doi [Science] {10.1126/science.1219631}, \href
  {http://adsabs.harvard.edu/abs/2012Sci...337..457I} {337, 457}

\bibitem[\protect\citeauthoryear{{Jacobson} et~al.,}{{Jacobson}
  et~al.}{2006}]{cassini_1}
{Jacobson} R.~A.,  et~al., 2006, \mn@doi [\aj] {10.1086/508812}, \href
  {https://ui.adsabs.harvard.edu/abs/2006AJ....132.2520J} {132, 2520}

\bibitem[\protect\citeauthoryear{{Jain} \& {Khoury}}{{Jain} \&
  {Khoury}}{2010}]{Jain_Khoury}
{Jain} B.,  {Khoury} J.,  2010, \mn@doi [Annals of Physics]
  {10.1016/j.aop.2010.04.002}, \href
  {https://ui.adsabs.harvard.edu/abs/2010AnPhy.325.1479J} {325, 1479}

\bibitem[\protect\citeauthoryear{Jeffreys}{Jeffreys}{1939}]{JS}
Jeffreys H.,  1939, {The Theory of Probability}.
Oxford Classic Texts in the Physical Sciences, Oxford, England

\bibitem[\protect\citeauthoryear{{Kent}}{{Kent}}{1987}]{Kent_a0}
{Kent} S.~M.,  1987, \mn@doi [\aj] {10.1086/114366}, \href
  {https://ui.adsabs.harvard.edu/abs/1987AJ.....93..816K} {93, 816}

\bibitem[\protect\citeauthoryear{{Kroupa} et~al.,}{{Kroupa}
  et~al.}{2018}]{Kroupa_2018}
{Kroupa} P.,  et~al., 2018, \mn@doi [\nat] {10.1038/s41586-018-0429-z}, \href
  {https://ui.adsabs.harvard.edu/abs/2018Natur.561E...4K} {561, E4}

\bibitem[\protect\citeauthoryear{{Kroupa} et~al.,}{{Kroupa}
  et~al.}{2022}]{Kroupa_2022}
{Kroupa} P.,  et~al., 2022, \mn@doi [\mnras] {10.1093/mnras/stac2563}, \href
  {https://ui.adsabs.harvard.edu/abs/2022MNRAS.517.3613K} {517, 3613}

\bibitem[\protect\citeauthoryear{{Kuzio de Naray} \& {Kaufmann}}{{Kuzio de
  Naray} \& {Kaufmann}}{2011}]{kuzio:2011}
{Kuzio de Naray} R.,  {Kaufmann} T.,  2011, \mn@doi [\mnras]
  {10.1111/j.1365-2966.2011.18656.x}, \href
  {https://ui.adsabs.harvard.edu/abs/2011MNRAS.414.3617K} {414, 3617}

\bibitem[\protect\citeauthoryear{{Lelli}, {McGaugh}  \& {Schombert}}{{Lelli}
  et~al.}{2016a}]{SPARC}
{Lelli} F.,  {McGaugh} S.~S.,   {Schombert} J.~M.,  2016a, \mn@doi [\aj]
  {10.3847/0004-6256/152/6/157}, \href
  {http://adsabs.harvard.edu/abs/2016AJ....152..157L} {152, 157}

\bibitem[\protect\citeauthoryear{{Lelli}, {McGaugh}  \& {Schombert}}{{Lelli}
  et~al.}{2016b}]{Federico_BTFR}
{Lelli} F.,  {McGaugh} S.~S.,   {Schombert} J.~M.,  2016b, \mn@doi [\apjl]
  {10.3847/2041-8205/816/1/L14}, \href
  {https://ui.adsabs.harvard.edu/abs/2016ApJ...816L..14L} {816, L14}

\bibitem[\protect\citeauthoryear{{Lelli}, {McGaugh}, {Schombert}  \&
  {Pawlowski}}{{Lelli} et~al.}{2017}]{OneLaw}
{Lelli} F.,  {McGaugh} S.~S.,  {Schombert} J.~M.,   {Pawlowski} M.~S.,  2017,
  \mn@doi [\apj] {10.3847/1538-4357/836/2/152}, \href
  {http://adsabs.harvard.edu/abs/2017ApJ...836..152L} {836, 152}

\bibitem[\protect\citeauthoryear{{Li} et~al.,}{{Li} et~al.}{2023}]{Li_clusters}
{Li} P.,  et~al., 2023, \mn@doi [\aap] {10.1051/0004-6361/202346431}, \href
  {https://ui.adsabs.harvard.edu/abs/2023A&A...677A..24L} {677, A24}

\bibitem[\protect\citeauthoryear{{Maquet} \& {Pierret}}{{Maquet} \&
  {Pierret}}{2015}]{maquet:2015jk}
{Maquet} L.,  {Pierret} F.,  2015, \mn@doi [\prd] {10.1103/PhysRevD.91.084015},
  \href {http://adsabs.harvard.edu/abs/2015PhRvD..91h4015M} {91, 084015}

\bibitem[\protect\citeauthoryear{{McGaugh} \& {Milgrom}}{{McGaugh} \&
  {Milgrom}}{2013}]{McGaugh_Milgrom}
{McGaugh} S.,  {Milgrom} M.,  2013, \mn@doi [\apj]
  {10.1088/0004-637X/775/2/139}, \href
  {https://ui.adsabs.harvard.edu/abs/2013ApJ...775..139M} {775, 139}

\bibitem[\protect\citeauthoryear{{McGaugh}, {Schombert}, {Bothun}  \& {de
  Blok}}{{McGaugh} et~al.}{2000}]{McGaugh_BTFR}
{McGaugh} S.~S.,  {Schombert} J.~M.,  {Bothun} G.~D.,   {de Blok} W.~J.~G.,
  2000, \mn@doi [\apjl] {10.1086/312628}, \href
  {https://ui.adsabs.harvard.edu/abs/2000ApJ...533L..99M} {533, L99}

\bibitem[\protect\citeauthoryear{{McGaugh}, {Lelli}  \& {Schombert}}{{McGaugh}
  et~al.}{2016}]{MLS}
{McGaugh} S.~S.,  {Lelli} F.,   {Schombert} J.~M.,  2016, \mn@doi [Physical
  Review Letters] {10.1103/PhysRevLett.117.201101}, \href
  {http://adsabs.harvard.edu/abs/2016PhRvL.117t1101M} {117, 201101}

\bibitem[\protect\citeauthoryear{McGaugh, Lelli  \& Schombert}{McGaugh
  et~al.}{2020}]{research_note}
McGaugh S.~S.,  Lelli F.,   Schombert J.~M.,  2020, \mn@doi [Research Notes of
  the AAS] {10.3847/2515-5172/ab8471}, 4, 45

\bibitem[\protect\citeauthoryear{{Milgrom}}{{Milgrom}}{1983a}]{Milgrom_1}
{Milgrom} M.,  1983a, \mn@doi [\apj] {10.1086/161130}, \href
  {http://adsabs.harvard.edu/abs/1983ApJ...270..365M} {270, 365}

\bibitem[\protect\citeauthoryear{{Milgrom}}{{Milgrom}}{1983b}]{Milgrom_3}
{Milgrom} M.,  1983b, \mn@doi [\apj] {10.1086/161131}, \href
  {http://adsabs.harvard.edu/abs/1983ApJ...270..371M} {270, 371}

\bibitem[\protect\citeauthoryear{{Milgrom}}{{Milgrom}}{1983c}]{Milgrom_2}
{Milgrom} M.,  1983c, \mn@doi [\apj] {10.1086/161132}, \href
  {http://adsabs.harvard.edu/abs/1983ApJ...270..384M} {270, 384}

\bibitem[\protect\citeauthoryear{{Milgrom}}{{Milgrom}}{1988}]{Milgrom_a0}
{Milgrom} M.,  1988, \mn@doi [\apj] {10.1086/166777}, \href
  {https://ui.adsabs.harvard.edu/abs/1988ApJ...333..689M} {333, 689}

\bibitem[\protect\citeauthoryear{{Milgrom}}{{Milgrom}}{1994}]{MI_1994}
{Milgrom} M.,  1994, \mn@doi [Annals of Physics] {10.1006/aphy.1994.1012},
  \href {https://ui.adsabs.harvard.edu/abs/1994AnPhy.229..384M} {229, 384}

\bibitem[\protect\citeauthoryear{{Milgrom}}{{Milgrom}}{2009}]{milgrom:2009vn}
{Milgrom} M.,  2009, \mn@doi [\mnras] {10.1111/j.1365-2966.2009.15302.x}, 399,
  474

\bibitem[\protect\citeauthoryear{{Milgrom}}{{Milgrom}}{2010}]{qumond}
{Milgrom} M.,  2010, \mn@doi [\mnras] {10.1111/j.1365-2966.2009.16184.x}, \href
  {https://ui.adsabs.harvard.edu/abs/2010MNRAS.403..886M} {403, 886}

\bibitem[\protect\citeauthoryear{Milgrom}{Milgrom}{2011}]{Milgrom_2011}
Milgrom M.,  2011, \mn@doi [Acta Phys. Polon. B] {10.5506/APhysPolB.42.2175},
  42, 2175

\bibitem[\protect\citeauthoryear{Milgrom}{Milgrom}{2022}]{inertia_1}
Milgrom M.,  2022, \mn@doi [Phys. Rev. D] {10.1103/PhysRevD.106.064060}, 106,
  064060

\bibitem[\protect\citeauthoryear{{Milgrom}}{{Milgrom}}{2023a}]{MI_2023}
{Milgrom} M.,  2023a, \mn@doi [arXiv e-prints] {10.48550/arXiv.2310.14334},
  \href {https://ui.adsabs.harvard.edu/abs/2023arXiv231014334M} {p.
  arXiv:2310.14334}

\bibitem[\protect\citeauthoryear{{Milgrom}}{{Milgrom}}{2023b}]{TRIMOND}
{Milgrom} M.,  2023b, \mn@doi [\prd] {10.1103/PhysRevD.108.063009}, \href
  {https://ui.adsabs.harvard.edu/abs/2023PhRvD.108f3009M} {108, 063009}

\bibitem[\protect\citeauthoryear{{Milgrom}}{{Milgrom}}{2023c}]{GQUMOND}
{Milgrom} M.,  2023c, \mn@doi [\prd] {10.1103/PhysRevD.108.084005}, \href
  {https://ui.adsabs.harvard.edu/abs/2023PhRvD.108h4005M} {108, 084005}

\bibitem[\protect\citeauthoryear{{Oman} et~al.,}{{Oman}
  et~al.}{2015}]{Oman_diversity}
{Oman} K.~A.,  et~al., 2015, \mn@doi [\mnras] {10.1093/mnras/stv1504}, \href
  {https://ui.adsabs.harvard.edu/abs/2015MNRAS.452.3650O} {452, 3650}

\bibitem[\protect\citeauthoryear{{Oria} et~al.,}{{Oria} et~al.}{2021}]{Oria}
{Oria} P.~A.,  et~al., 2021, \mn@doi [\apj] {10.3847/1538-4357/ac273d}, \href
  {https://ui.adsabs.harvard.edu/abs/2021ApJ...923...68O} {923, 68}

\bibitem[\protect\citeauthoryear{{Paranjape} \& {Sheth}}{{Paranjape} \&
  {Sheth}}{2021}]{Paranjape_Sheth}
{Paranjape} A.,  {Sheth} R.~K.,  2021, \mn@doi [\mnras]
  {10.1093/mnras/stab2141}, \href
  {https://ui.adsabs.harvard.edu/abs/2021MNRAS.507..632P} {507, 632}

\bibitem[\protect\citeauthoryear{Phan, Pradhan  \& Jankowiak}{Phan
  et~al.}{2019}]{phan2019composable}
Phan D.,  Pradhan N.,   Jankowiak M.,  2019, arXiv preprint arXiv:1912.11554

\bibitem[\protect\citeauthoryear{{Pittordis} \& {Sutherland}}{{Pittordis} \&
  {Sutherland}}{2023}]{Pittordis_Sutherland}
{Pittordis} C.,  {Sutherland} W.,  2023, \mn@doi [The Open Journal of
  Astrophysics] {10.21105/astro.2205.02846}, \href
  {https://ui.adsabs.harvard.edu/abs/2023OJAp....6E...4P} {6, 4}

\bibitem[\protect\citeauthoryear{{Pointecouteau} \& {Silk}}{{Pointecouteau} \&
  {Silk}}{2005}]{clusters_3}
{Pointecouteau} E.,  {Silk} J.,  2005, \mn@doi [\mnras]
  {10.1111/j.1365-2966.2005.09590.x}, \href
  {https://ui.adsabs.harvard.edu/abs/2005MNRAS.364..654P} {364, 654}

\bibitem[\protect\citeauthoryear{{Roper}, {Oman}, {Frenk},
  {Ben{\'\i}tez-Llambay}, {Navarro}  \& {Santos-Santos}}{{Roper}
  et~al.}{2023}]{Roper_diversity}
{Roper} F.~A.,  {Oman} K.~A.,  {Frenk} C.~S.,  {Ben{\'\i}tez-Llambay} A.,
  {Navarro} J.~F.,   {Santos-Santos} I. M.~E.,  2023, \mn@doi [\mnras]
  {10.1093/mnras/stad549}, \href
  {https://ui.adsabs.harvard.edu/abs/2023MNRAS.521.1316R} {521, 1316}

\bibitem[\protect\citeauthoryear{{Sanders}}{{Sanders}}{1999}]{clusters_2}
{Sanders} R.~H.,  1999, \mn@doi [\apjl] {10.1086/311865}, \href
  {https://ui.adsabs.harvard.edu/abs/1999ApJ...512L..23S} {512, L23}

\bibitem[\protect\citeauthoryear{{Schombert} \& {McGaugh}}{{Schombert} \&
  {McGaugh}}{2014}]{Schombert}
{Schombert} J.,  {McGaugh} S.,  2014, \mn@doi [\pasa] {10.1017/pasa.2014.32},
  \href {https://ui.adsabs.harvard.edu/abs/2014PASA...31...36S} {31, e036}

\bibitem[\protect\citeauthoryear{Schwarz}{Schwarz}{1978}]{BIC}
Schwarz G.,  1978, \mn@doi [The Annals of Statistics] {10.1214/aos/1176344136},
  6, 461

\bibitem[\protect\citeauthoryear{{Skordis} \& {Z{\l}o{\'s}nik}}{{Skordis} \&
  {Z{\l}o{\'s}nik}}{2021}]{skordis}
{Skordis} C.,  {Z{\l}o{\'s}nik} T.,  2021, \mn@doi [\prl]
  {10.1103/PhysRevLett.127.161302}, \href
  {https://ui.adsabs.harvard.edu/abs/2021PhRvL.127p1302S} {127, 161302}

\bibitem[\protect\citeauthoryear{{Stiskalek} \& {Desmond}}{{Stiskalek} \&
  {Desmond}}{2023}]{Stiskalek}
{Stiskalek} R.,  {Desmond} H.,  2023, \mn@doi [\mnras]
  {10.1093/mnras/stad2675}, \href
  {https://ui.adsabs.harvard.edu/abs/2023MNRAS.525.6130S} {525, 6130}

\bibitem[\protect\citeauthoryear{{Swaters}, {Sancisi}, {van Albada}  \& {van
  der Hulst}}{{Swaters} et~al.}{2009}]{swaters:2009}
{Swaters} R.~A.,  {Sancisi} R.,  {van Albada} T.~S.,   {van der Hulst} J.~M.,
  2009, \mn@doi [\aap] {10.1051/0004-6361:200810516}, \href
  {https://ui.adsabs.harvard.edu/abs/2009A&A...493..871S} {493, 871}

\bibitem[\protect\citeauthoryear{{The} \& {White}}{{The} \&
  {White}}{1988}]{clusters_1}
{The} L.~S.,  {White} S. D.~M.,  1988, \mn@doi [\aj] {10.1086/114760}, \href
  {https://ui.adsabs.harvard.edu/abs/1988AJ.....95.1642T} {95, 1642}

\bibitem[\protect\citeauthoryear{{Thomas}, {Famaey}, {Ibata}, {Renaud},
  {Martin}  \& {Kroupa}}{{Thomas} et~al.}{2018}]{Thomas}
{Thomas} G.~F.,  {Famaey} B.,  {Ibata} R.,  {Renaud} F.,  {Martin} N.~F.,
  {Kroupa} P.,  2018, \mn@doi [\aap] {10.1051/0004-6361/201731609}, \href
  {http://cdsads.u-strasbg.fr/abs/2018A%26A...609A..44T} {609, A44}

\bibitem[\protect\citeauthoryear{{Toomre}}{{Toomre}}{1963}]{Toomre}
{Toomre} A.,  1963, \mn@doi [\apj] {10.1086/147653}, \href
  {https://ui.adsabs.harvard.edu/abs/1963ApJ...138..385T} {138, 385}

\bibitem[\protect\citeauthoryear{{Tully} \& {Fisher}}{{Tully} \&
  {Fisher}}{1977}]{TFR}
{Tully} R.~B.,  {Fisher} J.~R.,  1977, \aap, \href
  {https://ui.adsabs.harvard.edu/abs/1977A&A....54..661T} {54, 661}

\bibitem[\protect\citeauthoryear{{Vokrouhlick{\'y}}, {Nesvorn{\'y}}  \&
  {Tremaine}}{{Vokrouhlick{\'y}} et~al.}{2024}]{2024:vokrouhlicky}
{Vokrouhlick{\'y}} D.,  {Nesvorn{\'y}} D.,   {Tremaine} S.,  2024, \mn@doi
  [arXiv e-prints] {10.48550/arXiv.2403.09555}, \href
  {https://ui.adsabs.harvard.edu/abs/2024arXiv240309555V} {p. arXiv:2403.09555}

\bibitem[\protect\citeauthoryear{Wolf \& Smith}{Wolf \&
  Smith}{1995}]{wolf:1995ys}
Wolf A.,  Smith J.,  1995, Control Engineering Practice, 3, 1611

\bibitem[\protect\citeauthoryear{{Wu}, {Famaey}, {Gentile}, {Perets}  \&
  {Zhao}}{{Wu} et~al.}{2008}]{Wu_2008}
{Wu} X.,  {Famaey} B.,  {Gentile} G.,  {Perets} H.,   {Zhao} H.,  2008, \mn@doi
  [\mnras] {10.1111/j.1365-2966.2008.13198.x}, \href
  {https://ui.adsabs.harvard.edu/abs/2008MNRAS.386.2199W} {386, 2199}

\bibitem[\protect\citeauthoryear{{Zhao} \& {Famaey}}{{Zhao} \&
  {Famaey}}{2006}]{zhao:2006la}
{Zhao} H.~S.,  {Famaey} B.,  2006, \mn@doi [\apjl] {10.1086/500805}, \href
  {http://adsabs.harvard.edu/abs/2006ApJ...638L...9Z} {638, L9}

\bibitem[\protect\citeauthoryear{Zhao \& Famaey}{Zhao \& Famaey}{2012}]{emond}
Zhao H.,  Famaey B.,  2012, \mn@doi [Phys. Rev. D]
  {10.1103/PhysRevD.86.067301}, 86, 067301

\bibitem[\protect\citeauthoryear{{Zwaan}, {van der Hulst}, {de Blok}  \&
  {McGaugh}}{{Zwaan} et~al.}{1995}]{zwaan:1995}
{Zwaan} M.~A.,  {van der Hulst} J.~M.,  {de Blok} W.~J.~G.,   {McGaugh} S.~S.,
  1995, \mn@doi [\mnras] {10.1093/mnras/273.1.L35}, \href
  {https://ui.adsabs.harvard.edu/abs/1995MNRAS.273L..35Z} {273, L35}

\bibitem[\protect\citeauthoryear{{de Blok} \& {McGaugh}}{{de Blok} \&
  {McGaugh}}{1997}]{deblok:1997}
{de Blok} W.~J.~G.,  {McGaugh} S.~S.,  1997, \mn@doi [\mnras]
  {10.1093/mnras/290.3.533}, \href
  {https://ui.adsabs.harvard.edu/abs/1997MNRAS.290..533D} {290, 533}

\makeatother
\end{thebibliography}

\label{lastpage}
\end{document}